# Dynamic traversal of large gaps by insects and legged robots reveals a template

Sean W. Gart, Changxin Yan, Ratan Othayoth, Zhiyi Ren, and Chen Li

Department of Mechanical Engineering, Johns Hopkins University

3400 N. Charles St, 126 Hackerman Hall, Baltimore, Maryland 21218-2683, USA

E-mail: [redacted]

## Abstract

It is well known that animals can use neural and sensory feedback via vision, tactile sensing, and echolocation to negotiate obstacles. Similarly, most robots use deliberate or reactive planning to avoid obstacles, which relies on prior knowledge or high-fidelity sensing of the environment. However, during dynamic locomotion in complex, novel, 3-D terrains such as forest floor and building rubble, sensing and planning suffer bandwidth limitation and large noise and are sometimes even impossible. Here, we study rapid locomotion over a large gap, a simple, ubiquitous obstacle, to begin to discover general principles of dynamic traversal of large 3-D obstacles. We challenged the discoid cockroach and an open-loop six-legged robot to traverse a large gap of varying length. Both the animal and the robot could dynamically traverse a gap as large as 1 body length by bridging the gap with its head, but traversal probability decreased with gap length. Based on these observations, we developed a template that well captured body dynamics and quantitatively predicted traversal performance. Our template revealed that high approach speed, initial body pitch, and initial body pitch angular velocity facilitated dynamic traversal, and successfully predicted a new strategy of using body pitch control that increased the robot's maximal traversal gap length by 50%. Our study established the first template of dynamic locomotion beyond planar surfaces and is an important step in expanding terradynamics into complex 3-D terrains.

# 1. Introduction

Complex 3-D terrains such as leaf litter, fallen branches on a forest floor, and landslide debris (figure 1) can pose a major challenge for small animals and robots alike, because obstacles are often comparable to, or even larger than, the animal or robot itself [1] and can induce large perturbations [2–8]. Despite this, small animals like insects, reptiles, and small birds agilely traverse complex 3-D terrains [2, 9] as diverse as inclined and vertical surfaces [10–16], thin ledges [17], large gaps [4, 7, 18–21] and bumps [3, 5, 22–26], uneven surfaces [6, 27–30], cluttered terrain [8], and even confined spaces [31], with locomotor performance far exceeding that of even the best robots today [32–36].

To move through varying complex environments, animals not only use multi-modal sensory feedback and feedforward neural commands to control and adjust their body and limbs, but their body and limbs also often have well-tuned morphology to accommodate perturbations in the environment via mechanical feedback [2, 37]. During slow locomotion where sensing is sufficient, animals like stick insects [4, 38], cockroaches [3, 23, 39, 40], lizards [5, 24], and snakes [7, 19] tend to use deliberate, seemingly well-planned, and often precisely-controlled, body and limb movements to negotiate with and traverse complex 3-D terrains, presumably using antennae or vision in the process. For example, stick insects use their antennae to sense the terrain and use quasi-static, “follow-the-leader” stepping to slowly walk on branches [41] climb over steps [11, 42] and bridge over large gaps . Slowly running [39, 43] or walking [3, 23, 40, 44] cockroaches use their antennae to sense obstacles in front and alter their kinematics to either climb over steps [3, 22], tunnel under steps [23, 45], approach and climb up pillars [44, 46], or follow walls [39, 43, 47] depending on the location of the obstacle. Lizards frequently jump onto and over large bumps [48], and snakes either quasi-statically cantilever their body to reach across a smaller gap [19] or dynamically lunge to traverse a larger gap [19], presumably all using vision in the process.

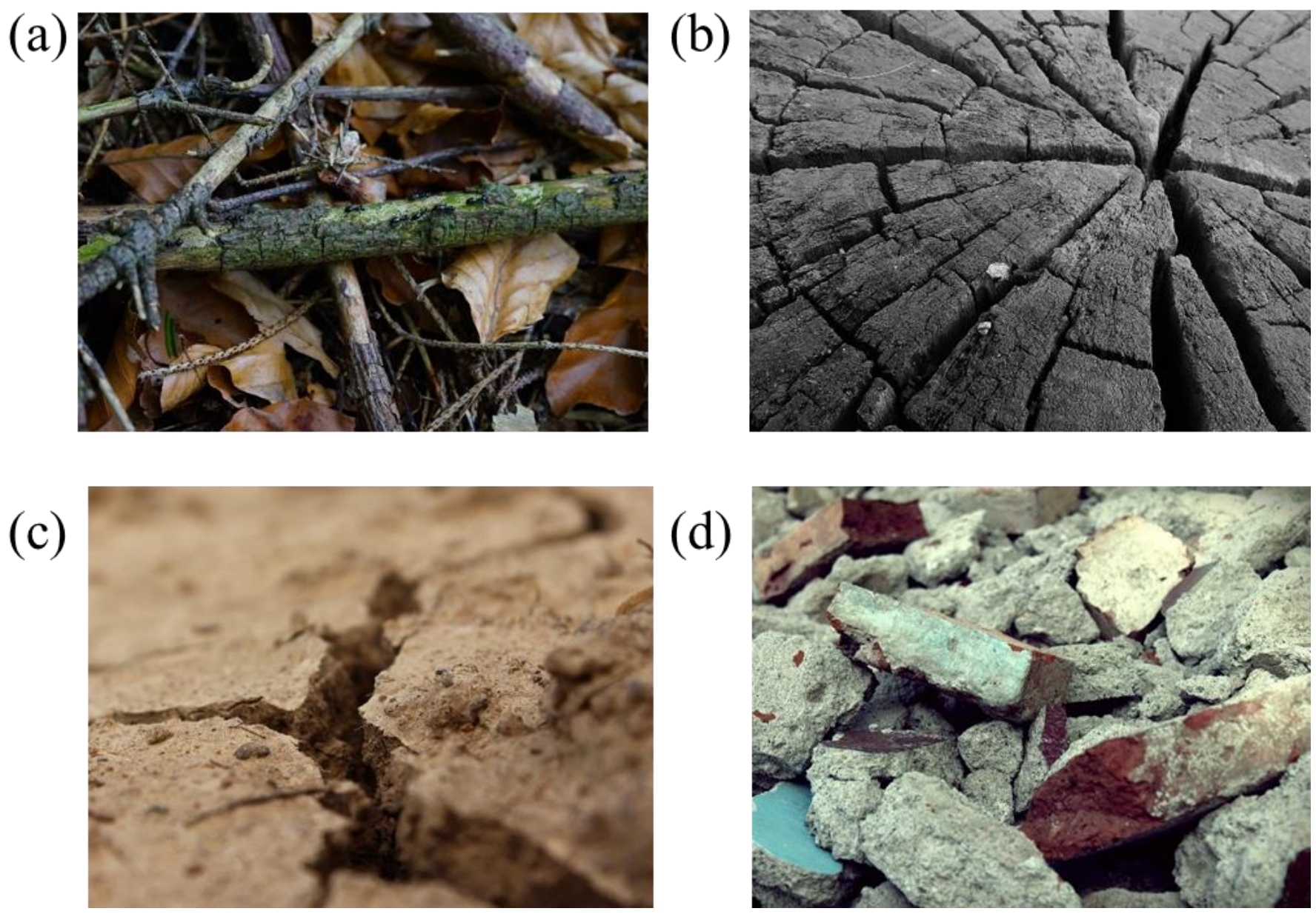

**Figure 1.** Examples of gap-like obstacles in complex terrains: (a) Leaf and branch litter, (b) tree crevices, (c) cracked dry soil, and (d) building rubble. All images were available under Creative Commons CC0 at pixabay.com.

By contrast, during rapid locomotion such as prey chasing and predator escape, particularly in terrains with frequent large disturbances [25], sensory noise [37] and sensory and neuromuscular delays [49] limit an animal's ability to plan its course of locomotion and control its body and limb movement during each step or gait cycle. To accommodate this, animals primarily move their body and limbs in a more feedforward manner [37] to simplify control, and use neural and sensory feedback to adjust body and limbs in response to large environmental perturbations. Because of this, how its body and limbs passively respond to the environment has a crucial impact on the dynamics and locomotor performance of a small animal moving dynamically through complex 3-D terrains [6, 8, 28, 50]. For example, cockroaches have streamlined body shapes that facilitate traversal of cluttered terrain , highly compressible yet robust exoskeletons that help them move through small crevices [31], sprawled leg posture that self-stabilizes lateral perturbations [51], viscoelastic leg muscles and tendons that dampen external perturbations [52, 53], and distributed leg spines that increase probability of firm foothold on low contact area surfaces [28].

Traditionally, the field of mobile robotics has mainly dealt with locomotion in complex environments by solving the problem of obstacle avoidance [54–56], which requires either prior knowledge or high-fidelity sensing to plan clear locomotion paths [56–58]. More recently, thanks to cross-discipline collaborations between biologists, applied mathematicians, and engineers, the neuromechanical principles from biological studies have enabled many under-actuated bio-inspired robots [32, 35, 59–61]. By combining high-level sensing and planning with mechanical feedback via mechanically tuned designs and control algorithms [62–64], these robots have achieved unprecedented locomotion performance on simple ground [32, 59] and are beginning to traverse complex 3-D terrains [8, 31, 36, 61].

For both legged animals and robots, primary focus of mechanical feedback studies has been how leg morphology and mechanics interact with simple ground to assist locomotion [37, 63, 65]. In complex 3-D terrains where obstacles could be larger than the animal or robot, the body of a small legged animal or robot, which is more robust to collision [31, 50, 66] may also physically contact and interact with the terrain to help locomotion. However, much less attention has been paid to the role of body-terrain interaction in legged locomotion until very recently [31, 50].

Here, we take the next step in understanding how body-terrain interaction affects dynamic traversal of large 3-D obstacles. We chose to focus on two simple large obstacles: (1) a gap obstacle as large as the animal / robot's body length, reported in this paper; and (2) a bump obstacle as high as four times the animal / robot's hip height, reported in a companion paper [67]. Such simple, well-defined, parameterizable laboratory models of natural terrains are useful towards understanding complex interactions between moving animals and their natural environment [3, 8, 10, 14, 68]. We chose to study the discoid cockroach, which lives on the floor of tropical rainforests and naturally traverses a wide variety of 3-D obstacles such as dense vegetation and litter [69], because mechanical feedback plays a major role in its locomotion at high speeds [6, 28, 37]. Better ability to traverse large gap and bump obstacles is also important for small legged robots in complex 3-D terrains like building rubble and landslides (figure 1(c)) during applications like search and rescue [70]. We chose to study a cockroach-inspired RHex-like robot [32] because it not only has similar body plan and running dynamics [71] as the animal and allows a direct comparison with animal observations, but also provides a physical model to precisely control and systematically vary locomotor parameters [70, 72].

In this paper, we focus on understanding how legged animal and robot's body dynamics and body-terrain interaction affects dynamic traversal of a large gap obstacle. We challenged the rapidly running animal and the open-loop robot with a large gap of variable size, and tested how the ability to traverse depended on gap length, running speeds towards the gap, and body orientation. Comparison of the animal with the open-loop robot provided insights into the role of active sensory feedback. Inspired by the similarities of animal and robot observations, we developed a reduced-order dynamic model, or template [10, 37, 73], to capture low-order dynamic traversal of a large gap obstacle. Finally, we added an active tail [21, 74] to the robot and tested a control strategy revealed by our template to enhance large gap traversal. In our companion paper [67], we report our experiments and modeling of dynamic traversal of a large bump obstacle, and discuss common and potentially general, principles and distinct differences between dynamic traversal of large gap and bump obstacles.

## 2. Methods

### 2.1. Animals

For animal experiments, we used seven male *Blaberus discoidalis* cockroaches (Pinellas County Reptiles, St Petersburg, FL, USA), as females were often gravid and under different load bearing conditions. Prior to experiments we kept the animals in individual plastic containers at room temperature (22 °C) on a 12h:12h light:dark cycle and added food (fish and rabbit pellets) *ad libitum*. Animals weighed 2.7 ± 0.2 g and measured 4.6 ± 0.2 cm in length, 2.3 ± 0.2 cm in width, and 0.7 ± 0.1 cm in thickness.

### 2.2. Legged robot

For robot experiments, we constructed a cockroach-inspired, six-legged robot by modifying the RHex-class design[37] (figure 2(b)). The slightly-flexible robot chassis was cut from a 3.1 mm thick acrylic sheet using a VLS 6.60 laser cutter (Universal Laser Systems Inc., Scottsdale, AZ). We attached each motor (460 RPM micro-gear DC motor, 50:1 gear ratio, ServoCity, Winfield, KS) that drove the legs to 2 cm × 2 cm pieces of 0.15 cm thick polystyrene that were subsequently attached to the chassis. To increase the maximal leg frequency given the motor used to drive the legs, we chose s-shape legs. The s-shape legs were 3D printed with PLA plastic (Ultimaker 2 Extended +, Ultimaker North America, Cambridge, MA). We wrapped each of the legs with friction tape (Duck Brand, Avon, OH) to increase ground traction. To generate stable spring-mass-like [75] running, we tuned the stiffness of the chassis and legs and leg friction. To approximate the anterior shape of the cockroach body for direct comparison of terrain interaction between the robot and the animal, we thermo-formed a bottom half of a quarter ellipsoid polystyrene shell, and attached it to the front of the robot. The robot measured 25 cm long, 16 cm wide, 8 cm tall, and weighed 194 g. We calculated robot moment of inertia by approximating the body as a simple rigid rod measuring 25 cm long and weighing 194 g, which rotates about a fixed end ($I = 1/3mb^2 = 0.004$ kg m$^2$, where $m$ is body mass and $b$ is body length).

To test the effect of passive mechanics, we did not implement any sensors and drove the robot with an open-loop leg control. We varied the robot's running speed from 50 cm s$^{-1}$ to 200 cm s$^{-1}$ by changing voltage supplied to the DC motors to adjust leg frequency. At the maximal voltage of 25 V, the robot ran stably at 8.0 ± 0.5 body lengths s$^{-1}$ (200 ± 12 cm s$^{-1}$).

### 2.3. Gap obstacle track

We constructed a 90 cm long, 30 cm wide track with a gap that spanned the entire width of the track (figure 2(c)) using t-slotted extruded aluminum and acrylic (McMaster-Carr, Elmhurst, IL, USA). For both animal and robot experiments, we varied gap length $L$ from 0.2 body length to 1 body length ($b \approx 5$ cm for the animal, $b = 25$ cm for the robot) by sliding the far side of the track. To measure the maximal traversable gap length, we made an infinite length gap (or cliff) by removing the far side of the gap. We covered the entire test surface with white paper cardstock (Pacon 4-ply railroad poster board, Appleton, WI, USA) for animal experiments. For robot experiments, we covered the near side of the test surface with 50 grit sandpaper to increase leg traction and prevent slipping when the legs started moving. To prevent scratching of the robot shell during impact, we covered the opposite side of the gap with polystyrene.

### 2.4. Experiment protocol

We filmed the animal and robot running over the gap at 500 frames $s^{-1}$ using three synchronized high-speed cameras, two from the dorsal view and one from the side view (figure 2(c)), with a shutter time of 500 µs. Dorsal cameras were placed directly above the near edge of the gap. A small lens aperture was used to maximize the focal depth of field.

We illuminated and heated the test area to 35 °C with 500 W work lamps (Coleman Cable, Waukegan, IL, USA), three from the dorsal side, and two laterally. To track the animal and robot, we attached a BEEtag [76], printed on standard office paper and attached to cardstock using double-sided tape, to the dorsal side of the body.

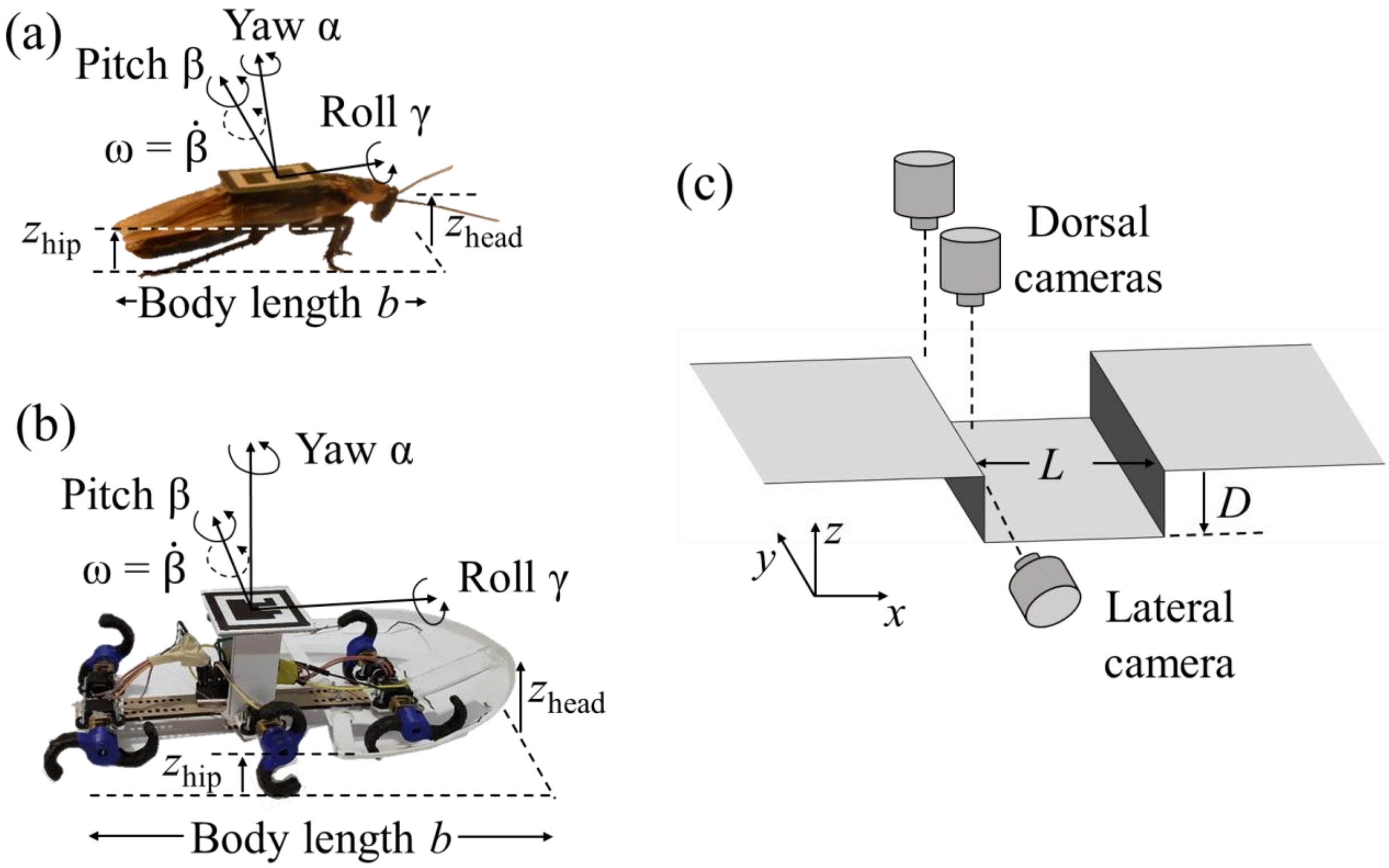

**Figure 2.** Experimental setup and definition of geometric and kinematic parameters and variables. (a) Discoid cockroach, body length $b$ = 4.6 cm ± 0.2 cm, hip height $z_{hip}$ = 0.5 cm. (b) Legged robot, body length $b$ = 25 cm, hip height $z_{hip}$ = 2.5 cm. BEEtags [76] were used to measure 3-D position ($x$, $y$, $z$), orientation (yaw α, pitch β, and roll γ), and pitch angular velocity ω of the body. (c) Schematic of the gap obstacle. Gap length $L$ was varied from 0.2 to 1 body length for both animal and robot experiments. Gap depth was fixed ($D$ = 0.7 body length for animal; $D$ = 0.3 body length for robot). Two dorsal and one lateral high speed cameras were used to record locomotion.

*2.4.1. Animal experiments*

We filmed animal experiments using Photron Mini UX100 cameras (Photron USA, San Diego, CA, USA) with a resolution of 1280 × 1024 pixels. For animal tracking, we attached a 1.6 cm × 1.6 cm BEEtag [76] to the dorsal surface of the wings directly above the body center of mass (CoM) [77] (figure 2(a)). We attached the tag to the animal using a combination of hot glue, super glue, and baking powder (as an accelerant). The animals were allowed to rest for at least 1 hour after tags were attached.

We placed cockroaches on the track one at a time for testing. To elicit a rapid escape response, we prodded the posterior and abdomen of the animal with a rod wrapped in paper tape. The animal ran towards the obstacle between two walls that funneled it towards the middle of the track. To encourage the animal to seek shelter [23, 45, 78], we placed a shaded overhang after the obstacle. The animals were allowed to rest for 1 to 2 minutes before each trial.

*2.4.2. Robot experiments*

We filmed all robot experiments using Fastec IL5 cameras (Fastec Imaging, San Diego, CA, USA) with a resolution of 1920 × 1080 pixels. For robot tracking with no tail, we attached a 5 cm × 5 cm tag above the robot body CoM using 0.3 cm thick polystyrene. The polystyrene was robust in cases where the robot flipped over. For experiments with a tail, we placed an additional 5 cm × 5 cm tag on the front shell of the robot, and the body CoM tag was moved to the side of the tail servo motor so that it was not obstructed by the tail when actuated. Three 3.8 cm × 3.8 cm tags were added to the dorsal side, ventral side, and tip of the tail to track tail motion.

*2.4.3. Tailed robot experiment*

To study how initial body pitch and initial body pitch angular velocity affected gap traversal, we added an active tail to the posterior end of the robot. We fastened the base of the 12 cm long active tail (3-D printed from PLA plastic) to a high torque servo motor (Futaba BLS274SV, Futaba, Champaign, IL) and attached a 33 g mass at its distal end. The active tail rotated within the sagittal plane at a maximal angular speed of 315 ± 115 ° $s^{-1}$. We tuned the base position (6 cm from the posterior end) and actuation timing (140 ± 14 ms prior to reaching the obstacle) of the active tail so that the robot's body pitch increased from −4° ± 8° prior to actuation to 8° ± 6° after actuation. The servo motor actuation was controlled by an Arduino Uno micro-controller and a motor controller (Qunqi L298 Dual H-bridge motor driver module). For this experiment, the robot legs were still under open-loop control. Adding the tail increased the robot's total body inertia along the pitch axis about its posterior end from 0.004 kg $m^2$ to 0.007 kg $m^2$ and did not change the robot's maximal running speed ($P$ = 0.4, Student's t-test).

To test the effect of body pitching, we ran the tailed robot perpendicular to the gap at a constant speed of 190 ± 16 cm $s^{-1}$ and varied whether the tail was activated or not for each gap length. We empirically timed tail activation so that the robot started to pitch up as the head reached the near edge of the gap. For finite length gap experiments, we increased gap length until the robot failed to traverse in all trials and decreased it until the robot successfully traversed in all trials. We collected 10 trials for each gap length tested. This resulted in a total of 40 trials for the experiment without tail actuation for a gap of 0.6, 0.8, 1.0, and infinite body length, and a total of 50 trials for the experiment with tail actuation for a gap of 0.8, 1.0, 1.2 1.4, and infinite body length ($N$ = 1, $n$ = 90).

*2.4.4. Kinematic Tracking*

To calibrate the cameras, we made a calibration grid with Lego bricks (The Lego Group, Bilund, Denmark) and placed it in view of all cameras prior to each experiment session. We obtained intrinsic (focal length and lens distortion) and extrinsic (relative positions and rotations) camera parameters using direct linear transformation (DLT) 3-D reconstruction [79]. After experiments, we imported image sequences into a custom MATLAB script that tracked the tags in each camera view using the BEEtag code [76]. With an additional custom DLT script [79], we obtained the 3-D position ($x$, $y$, $z$) and orientation (Euler angles yaw $\alpha$, pitch $\beta$, and roll $\gamma$) of the tags (figures 2(a, b)). To verify BEEtag accuracy, we 3-D printed a calibration object with 9 BEEtags equally spaced 7 cm apart in a grid in the horizontal plane, but orientated at pitch and yaw angles from 0° to 60° in an increment of 30°. We then calculated error by measuring the 3-D position and orientation all the tags and comparing them with the designed values. We found that BEEtags accurately measured body orientation angles (s.d. of error = 1.1°) and fore-aft position and body height (s.d. of error = 1.2 mm). We measured the position and orientation of the CoM relative to the tag and inferred the CoM for each animal. To examine how the animal and robot's head interacted with the obstacle, we calculated head position from the tag position, assuming that the body and head together acted as a rigid body. This was a reasonable approximation: our manual tracking verified that the head moved little (< 0.13 cm) relative to the tag throughout each running cycle, and only moved slightly longer (0.27 cm) when the animal flexed its neck joint and abdomen while gripping and climbing onto the far side of the gap. For the robot, we generated a 3-D point cloud of the shell using a CAD model (Solidworks, Solidworks corporation, Waltham, MA, USA) to obtain head position (anterior end of the shell) relative to tag position.

**2.5. Data analysis**

For all animal and robot experiments, we measured 3-D kinematics including speed $v$, body pitch $\beta$, body pitch angular velocity $\omega$, body yaw $\alpha$, and body roll $\gamma$ as a function of forward displacement $x$ and traversal probability. We categorized a trial as successful traversal if the animal and robot's CoM reached the far side of the gap, and as failure to traverse if any part of the body touched the bottom of the gap. For the infinite length gap experiment, we also measured the distance that the animal and robot's head reached before it fell below the sides of the gap (maximal traversable gap length, $d$).

For all experiments, we observed only a small difference between the velocity heading and body yaw immediately prior to gap encounter (7° ± 8° for the animal, 2° ± 8° for the robot). Therefore, we assumed that velocity heading always equaled body yaw. We defined angle of incidence $\theta_0$ as the angle between velocity heading and the forward $+x$ direction at the time of obstacle encounter, which equaled body yaw $\alpha_0$ at the time of obstacle encounter (see figure 2(a, b)). We calculated angle of incidence $\theta_0$, initial body pitch $\beta_0$, initial body pitch angular velocity $\omega_0$, initial body roll $\gamma_0$, and approach speed $v_0=v\cos\theta_0$ (speed perpendicular to the gap) immediately before the animal and robot encountered the near edge of the gap obstacle. Our use of approach speed accounted for any motion not perpendicular to the gap. All metrics were averaged over 4 frames to reduce instantaneous measurement error, except for initial body pitch angular velocity averaged from when the head reached until the body CoM passed the near edge of the gap to account for the large noise. We reported both body yaw $\alpha$ and angle of incidence $\theta_0$ in absolute values due to lateral symmetry.

To determine whether the animal and robot approached and traversed the gap obstacle at speeds comparable to that during walking or running, we used the Froude number [80] $Fr = v_0(gz_{\text{hip}})^{-1/2}$, where $g$ is acceleration due to gravity, $v_0$ is approach speed, and $z_{\text{hip}}$ is hip height (0.5 cm for the animal, 2.5 cm for the robot). As a common form of normalized speed of terrestrial locomotion, Froude number is a good predictor of the speed at which legged animals transition from a walking to a running gait [80]. For the discoid cockroach, this transition occurs at normalized speeds of $Fr$ = 1.5 to 1.7 [6, 71]. In the remainder the paper, we refer the locomotion speed perpendicular to the gap as approach speed.

**2.6. Statistics**

For animal gap experiments, seven animals ran 10 trials each over six different gap lengths of 0.2, 0.4, 0.6, 0.8, 1, and infinite body length, resulting in a total of 420 accepted trials ($N$ = 7, $n$ = 420). We rejected a trial whenever the animal collided with the wall, turned back, or stopped moving forward before encountering the gap, or if it moved out of the camera view during a traversal attempt. We varied gap length but allowed the animal to run with its own chosen speed and velocity heading during each trial.

For the robot gap experiment, we systematically varied gap length from 0.2 to 1 body length in increments of 0.2 body length and tested five different running speeds of 78 ± 15 cm $s^{-1}$, 120 ± 17 cm $s^{-1}$, 170 ± 18 cm $s^{-1}$, 190 ± 10 cm $s^{-1}$, and 200 ± 12 cm $s^{-1}$. The robot initial velocity heading was always perpendicular to the gap. We collected 5 trials for each combination of gap length and speed, resulting in a total of 110 trials ($n$ = 1, $N$ = 110). For the infinite length gap experiment, we tested five different approach speeds of 56 ± 13 cm $s^{-1}$, 78 ± 3.0 cm $s^{-1}$, 130 ± 11 cm $s^{-1}$, 180 ± 7.6 cm $s^{-1}$, and 200 ± 6.2 cm $s^{-1}$ and collected 5 trials at each speed for a total of 25 trials ($n$ = 1, $N$ = 25). With no tail, the robot had a small, nearly constant initial body pitch of 1° ± 3° (mean ± s.d.) immediately before reaching the near edge of the gap.

To compare metrics (approach speed, traversal probability, etc.) across a gap of different lengths, we first calculated the mean for each individual for each gap, and then averaged over individual means to obtain the cross-individual average for that gap. Because we used only one robot, we averaged all trials for each gap to obtain the mean. To test which metrics (approach speed, initial body pitch, etc.) affected traversal success or failure, we used multiple logistic regression. To test all other metrics reported, we used repeated-measures ANOVA for animal experiments and ANOVA for robot experiments. Our multiple logistic regression and repeated-measures ANOVA accounted for individual variance by using the individual as one of the factors. We used Tukey's honestly significant difference (HSD) method for post-hoc analysis where needed. All data are reported as means ± 1 standard deviation (s.d.) unless otherwise specified.

## 3. Results and discussion

### 3.1. Dynamic gap traversal performance

In our experiments, both the discoid cockroach and the robot dynamically ran at high speeds to traverse the gap obstacle. Animal running speeds were 61 ± 16 cm $s^{-1}$ (12 ± 4 body length $s^{-1}$), up to 9 times that in previous experiments of slow, quasi-static gap traversal [3, 16, 23]. The majority (83%) of animal trials had a Froude number above that for walking-to-running transition ($Fr = 1.5$) [6, 71]. Robot running speeds were 146 ± 48 m/s (6 ± 2 body length $s^{-1}$).

During such high speed locomotion, both the animal and the robot were able to dynamically traverse a large gap of up to 1 body length. Probability of dynamic traversal decreased with gap length (figure 5(a, b), solid curve; animal: $P < 0.0001$, multiple logistic regression; robot: $P < 0.0001$, multiple logistic regression). When the gap is small enough (up to 0.4 body length), both the animal and the robot almost always traversed (93% for the animal and 100% for the robot, respectively). By contrast, for the largest gap of 1 body length tested, traversal was unlikely for both the animal and the robot. Only two out of the seven cockroaches were able to traverse it, while the rest always failed to traverse; the robot was never able to traverse it at its highest running speed (8.0 ± 0.5 body lengths $s^{-1}$). We verified that no animals could traverse a gap larger than 1 body length.

**3.2. Gap bridging by head well predicts dynamic gap traversal**

Careful observation of how the animal and robot's body interacted with the gap revealed that dynamic traversal of a gap obstacle was well predicted by whether the head was able to bridge across the gap. As the animal and robot encountered a large gap, each pair of legs sequentially lost contact with the near side, and the body began falling, with the head pitching downward (figure 3(a, b, e, f), frames 1, 2, and 3; supplementary video 1, 2). Gap bridging by the head occurred when the animal or robot's head reached the far edge of the gap before falling below it (figure 3(a, e); figure 3(c, d, g, h), red solid curve). When gap bridging by the head did not occur, the animal and robot always fell into the gap and failed to traverse (figure 3(b, f); figure 3(c, d, g, h), blue dashed curves). We found that when gap bridging by the head occurred, traversal was successful 97% of the time for the animal and 81% of the time for the robot (the exceptional cases were due to failure to grip, discussed in Section 3.6). When the head did not bridge the gap, both the animal and the robot failed to traverse 100% of the time. For a small enough gap (below 0.6 body length), the animal and robot always traversed because the head could not fall below the surface over such a small forward displacement (figure S4, S5), and gap bridging was guaranteed.

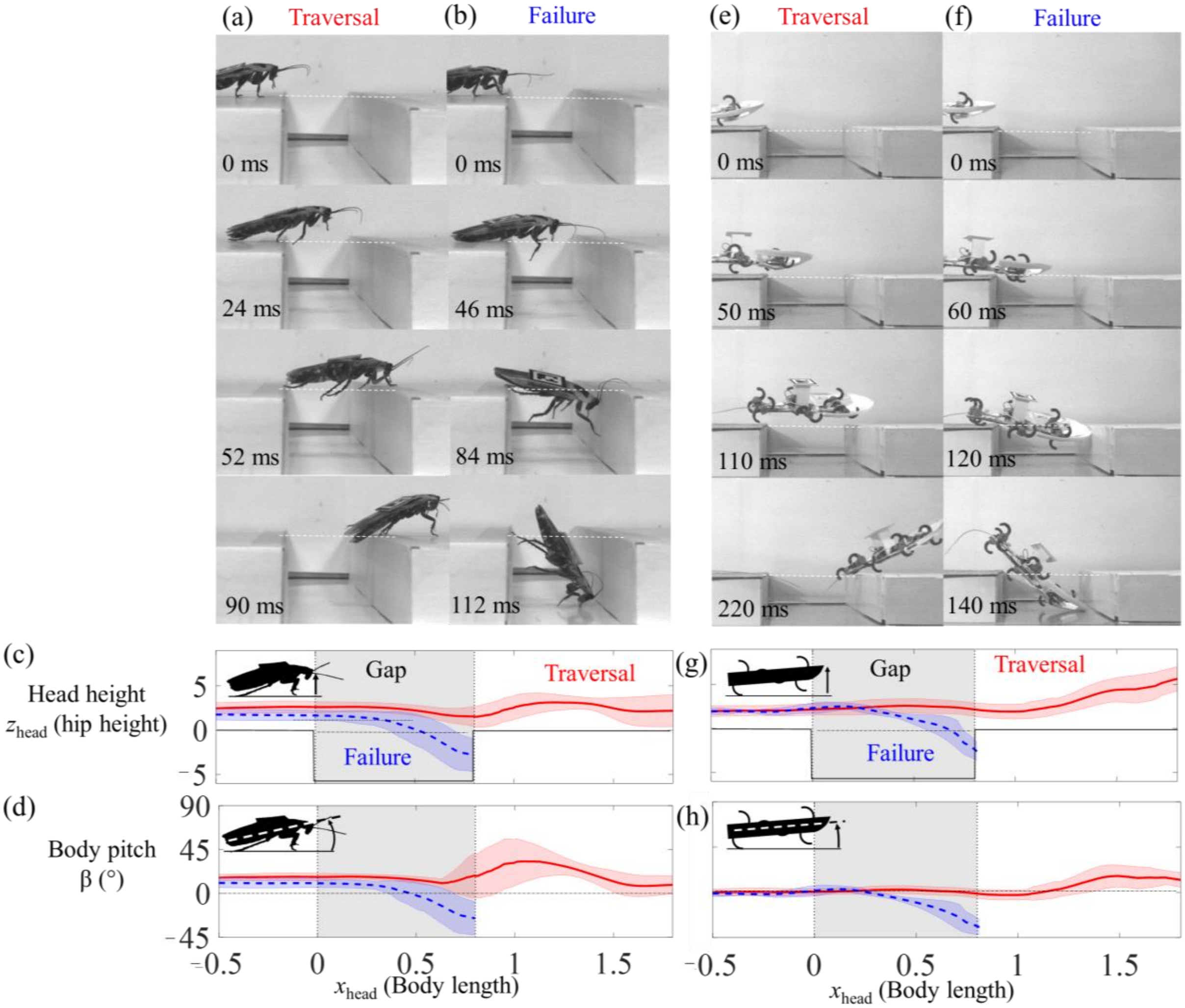

**Figure 4.** Dynamic locomotion of the discoid cockroach and the robot over a large gap obstacle. Representative trials of (a, e) successful traversal and (b, f) failure to bridge the gap. (c, g) Head height as a function of forward position of the head.* (d, h) Body pitch as a function of forward position of the head. In (c, d, g, h), solid red and blue dashed curves and shaded areas represent means ± 1 s.d. for the cases of success and failure. Data are shown for the 0.8 body length (4 cm) gap as an example; data for other gap lengths have similar trends (figures S4, S5, S6, S7). For simplicity, only movement perpendicular to the gap (within the *x-z* plane) is shown.

* We noted that when the animal failed to traverse, it often rebounded backwards after impacting the far side of the gap, but we could not track this motion due to the body and gap ledge obscuring the tag. Therefore, we truncated the data at the far edge of the gap.

The animal's gap bridging using the head was likely a passive process, similar to our robot under open-loop control. Although cockroaches actively sweep their antenna to sense the physical environment during slow exploration [40], we observed that the animal's antenna were held straight forward and slightly upward [40] and rarely came in contact with the ground. In addition, at the high speeds in our experiments, the animal had a very short time to respond if it did detect the gap (given a sensory delay of 6-40 ms [45] and a neuromuscular delay of 47 ms [6]), comparable to the time (60 ± 10 ms) for the animal's head to fall and bridge across the largest gap. Further, we observed that the animal rarely changed speed (by 3 ± 4 cm $s^{-1}$), body pitch (by 2° ± 3°), or heading (by 6° ± 6°) before reaching the gap. Together, these observations indicated that the animal was likely unable to respond in time. By contrast, later during dynamic gap traversal, the animal likely used sensory feedback to initiate and control active body and leg adjustments (see section 3.6).

**3.3. Template for dynamic gap traversal**

The striking similarities in traversal performance of the animal and the robot and the signature observation of gap bridging by the head for successful traversal suggested a template for dynamic gap traversal. A template is a useful modeling concept that allows general, fundamental understanding of high-dimensional, nonlinear, multi-body dynamic locomotion phenomena by reducing the problem to as few degrees of freedom as possible [37, 75]. A few templates have already captured fundamental dynamics and revealed control strategies for common forms of terrestrial locomotion on 2-D surfaces such as dynamic walking [81, 82] and running on level ground [62, 63, 83] and dynamic climbing on vertical walls [10, 15]. Inspired by these successes, we take the next step in creating a template for dynamic traversal of a large gap as a representative of complex 3-D terrains.

We approximated the animal and robot body as a rigid ellipsoid traveling forward at a constant approach speed, and calculated its dynamics during passive falling under gravity as it encountered a large gap (figure 4(a)). For simplicity, we assumed that the rigid body only rotated about the body pitch axis in the sagittal plane and had no body yaw or roll rotations. In addition, we assumed a fixed axis of rotation at the posterior end of the body at a height equaling the hip height of the animal (0.5 cm) or robot (2.5 cm). Finally, to model the body gradually losing support as legs gradually lost contact when the animal or robot ran past the near edge, we assumed that the force due to gravity increased proportionally to the length of the body beyond the near edge.

Using the Lagrangian method, we obtained the equation of motion during the pre-free falling phase (before the posterior passed the near edge) as:

$$\frac{\mathrm{d}}{\mathrm{d}t}\left(\frac{\partial L}{\partial \omega}\right) - \frac{\partial L}{\partial \beta}=0 \tag{1}$$

with the Lagrangian

$$L=\frac{1}{2}I\omega^2 - \frac{mgb}{2}\sin\beta \tag{2}$$

where β is body pitch, $I = 1/3\ mb^2$ is the moment of inertia for a body pitching about the hinge on its end, $b$ is the length of the body (5 cm for the animal, 25 cm for the robot), $m$ is body mass (2.5 g for the animal, 194 g for the robot), ω is the pitch angular velocity, and $g = 9.81$ m s$^{-2}$ is the gravitational acceleration. From the equation of motion, we obtained angular acceleration about the body pitch axis:

$$\dot{\omega}(t)= \frac{\xi mgb}{2I}\cos\beta(t) \tag{3}$$

where $\xi=\left(\frac{vt}{b}\right)\frac{\cos\beta(t)}{\cos\theta_0}$ is a linear force reduction factor that characterizes the loss of body support due to the legs losing surface contact, $v$ is the speed along the long axis of the body, and $\theta_0$ is the angle of incidence that measures the deviation of body yaw and velocity heading from the forward direction (direction perpendicular to the gap).

Once the posterior end of the body reached the near edge of the gap, the body started to fall like a projectile with zero angular acceleration and $\omega(t) = \omega_{edge}$, where $\omega_{edge}$ is the angular velocity of the body when the posterior reached the near gap edge. The center of mass height during the free falling phase is:

$$z_{\mathrm{CoM}}(\hat{t}) = \frac{1}{2} g \hat{t}^2, \quad (6)$$

$$\dot{\omega}\,(\hat{t}) = 0 \quad (7)$$

where $\hat{t}$ is the time after the posterior end passed the near edge of the gap. Note that we assumed that velocity heading and body yaw were always aligned based on experimental observations (see Section 2.5).

Finally, we used the Euler method to integrate forward in time to obtain body pitch angular velocity $\omega$, body pitch $\beta$, the center of mass position $x_{\mathrm{CoM}}$ along the forward direction, and the vertical center of mass position $z_{\mathrm{CoM}}$ as a function time $t$:

$$\omega(t+\Delta t) = \omega(t) + \dot{\omega}(t)\Delta t \quad (4)$$

$$\beta(t+\Delta t) = \beta(t) + \beta(t)\Delta t \quad (5)$$

$$x_{\mathrm{CoM}}(t+\Delta t) = x_{\mathrm{CoM}}(t) + \dot{x}_{\mathrm{CoM}}(t)\Delta t \quad (6)$$

$$z_{\mathrm{CoM}}(t+\Delta t) = z_{\mathrm{CoM}}(t) + \dot{z}_{\mathrm{CoM}}(t)\Delta t \quad (7)$$

where time step $\Delta t = 0.001$ seconds.

Assuming that gap bridging by the head results in traversal, the template allowed us to predict the maximal traversable gap length that the body could traverse for any given initial body pitch $\beta_0$, initial body pitch angular velocity $\omega_0$, and approach speed $v_0$ (supplementary video 4). We first numerically calculated time that the body had to attempt to bridge the gap, $t_{\mathrm{bridge}}$, i.e., the time for the rigid body's head (free anterior end) to fall to surface level (zero height). The forward distance over which the head travelled, $d = v_0 t_{\mathrm{fall}}$, was then the maximal traversable gap length (figure 4(b)).

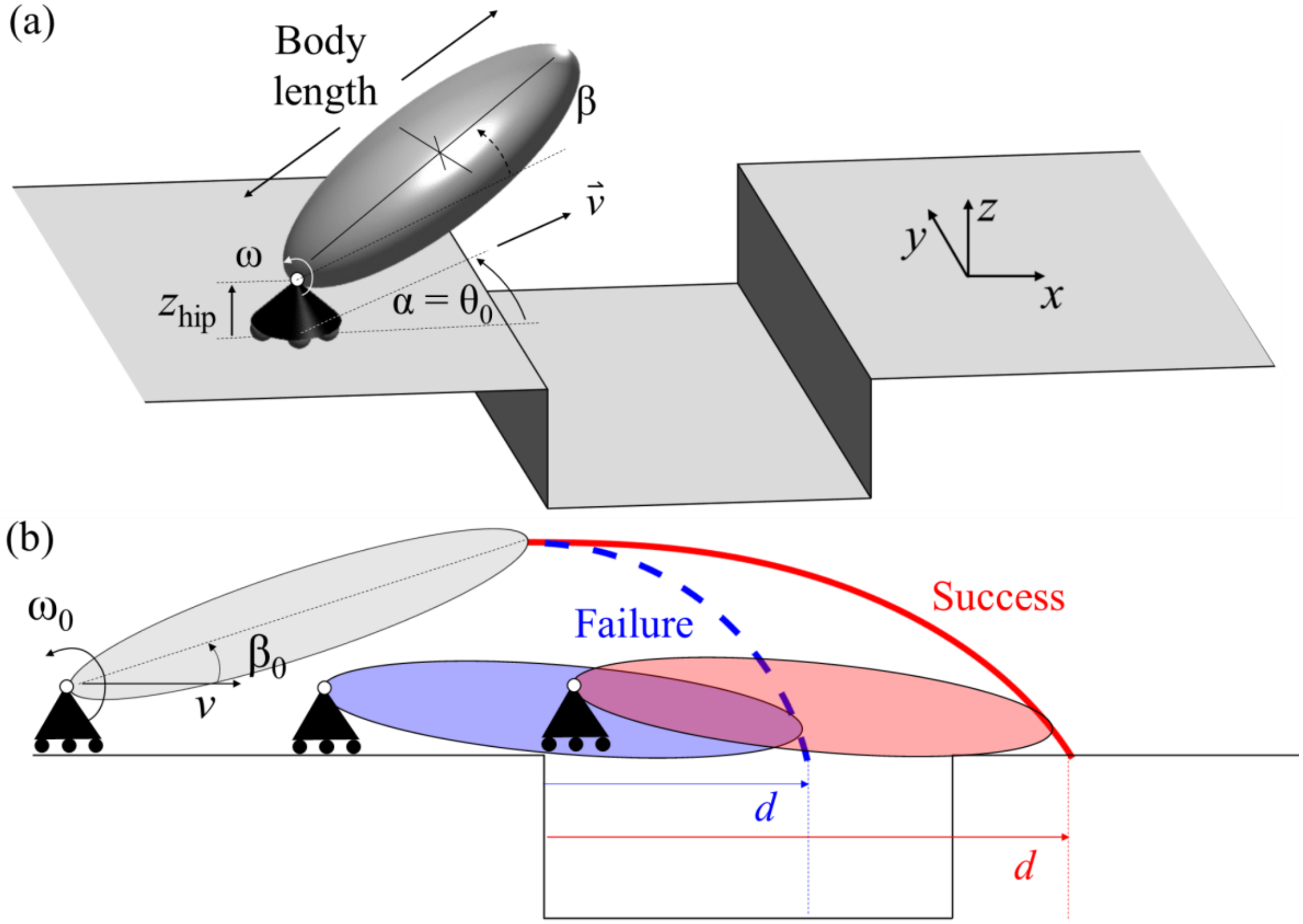

**Figure 4.** Template for dynamic gap traversal. (a) An animal or robot's body is modeled as a rigid body traveling at a constant speed $v$ along its body long axis, with a body pitch angle β and body pitch angular velocity ω. As it encounters a gap at an angle of incidence $\theta_0$, the body rotates downward in the sagittal plane about a hinge at its posterior end with a height equaling the animal or robot's hip height $z_{hip}$. The body yaw and velocity heading are assumed to be the same and at an angle α relative to the direction perpendicular to the gap. (b) Side view of model-predicted representative head trajectories and maximal traversable gap length, $d$, given the initial conditions $\omega_0 = 0°$ s$^{-1}$, $\beta_0 = 20°$, and a higher speed $v_0 = 65$ cm s$^{-1}$ (red solid curve), bridging the gap, and a lower speed $v_0 = 30$ cm s$^{-1}$ (blue dashed curve), falling into the gap.

To validate the template, we ran the animal and the robot into an infinite gap (a cliff) to measure maximal traversable gap length and compared it with model predictions. Without any model fitting parameters, the template well predicted maximal traversable gap length $d$ for both the animal and the robot (figure 5(a, b)). We noted that the animal and the robot ran at a broad range of approach speed (animal: 37 cm s$^{-1}$ $\leq v_0 \leq$ 100 cm s$^{-1}$; robot: 39 cm s$^{-1}$ $\leq v_0 \leq$ 210 cm s$^{-1}$), initial body pitch (animal: $3° \leq \beta_0 \leq 24°$; robot:

$-5° \le \beta_0 \le 9°$), and initial body pitch angular velocity (animal: $-401°\ s^{-1} \le \beta_0 \le 34°\ s^{-1}$; robot: $-143°\ s^{-1} \le \beta_0 \le 103°\ s^{-1}$) in the experiment.

The template slightly under-predicted maximal traversable gap length for the robot. One reason was due to its bouncier running gait, which resulted in a significant initial upward speed (20 ± 20 % of its forward speed, vs. the animal's 5 ± 4 %), not accounted for by the horizontally approaching template. Additionally, because the robot's head was in front of its legs by a large distance (0.3 body length), it did not start to lose ground reaction force and fall into the gap until it had already entered the gap substantially. By contrast, in the model, the body started to lose ground reaction force and fall immediately upon reaching the gap.

Next, we tested the predictive power of the template for predicting traversal probability measured in finite length gap experiments. Using the measured approach speed $v_0 = v\cos\theta_0$, initial body pitch $\beta_0$, and initial body pitch angular velocity $\omega_0$ from each experimental trial as initial conditions, we calculated the expected maximal traversable gap length from the model. If it exceeded the gap length being tested, the model-predicted successful traversal (figure 4(b), red solid curve); otherwise, the model-predicted failure (figure 4(b), blue dashed curve). Without any model fitting parameters, the template well predicted the observed monotonic decrease of traversal probability with gap length for both the animal and the robot (figure 5(c, d), dashed curves). For small gap length, quantitative agreement between animal data and model predictions was excellent. The over-prediction of traversal probability for the two largest gaps was because the model did not account for the animal or robot failing to grip the far side of the gap after the head bridged (see section 3.6). The template under-predicted robot traversal probability for the 0.4 and 0.6 body length gaps due to its bouncier gait and its leg substantially behind the head discussed above.

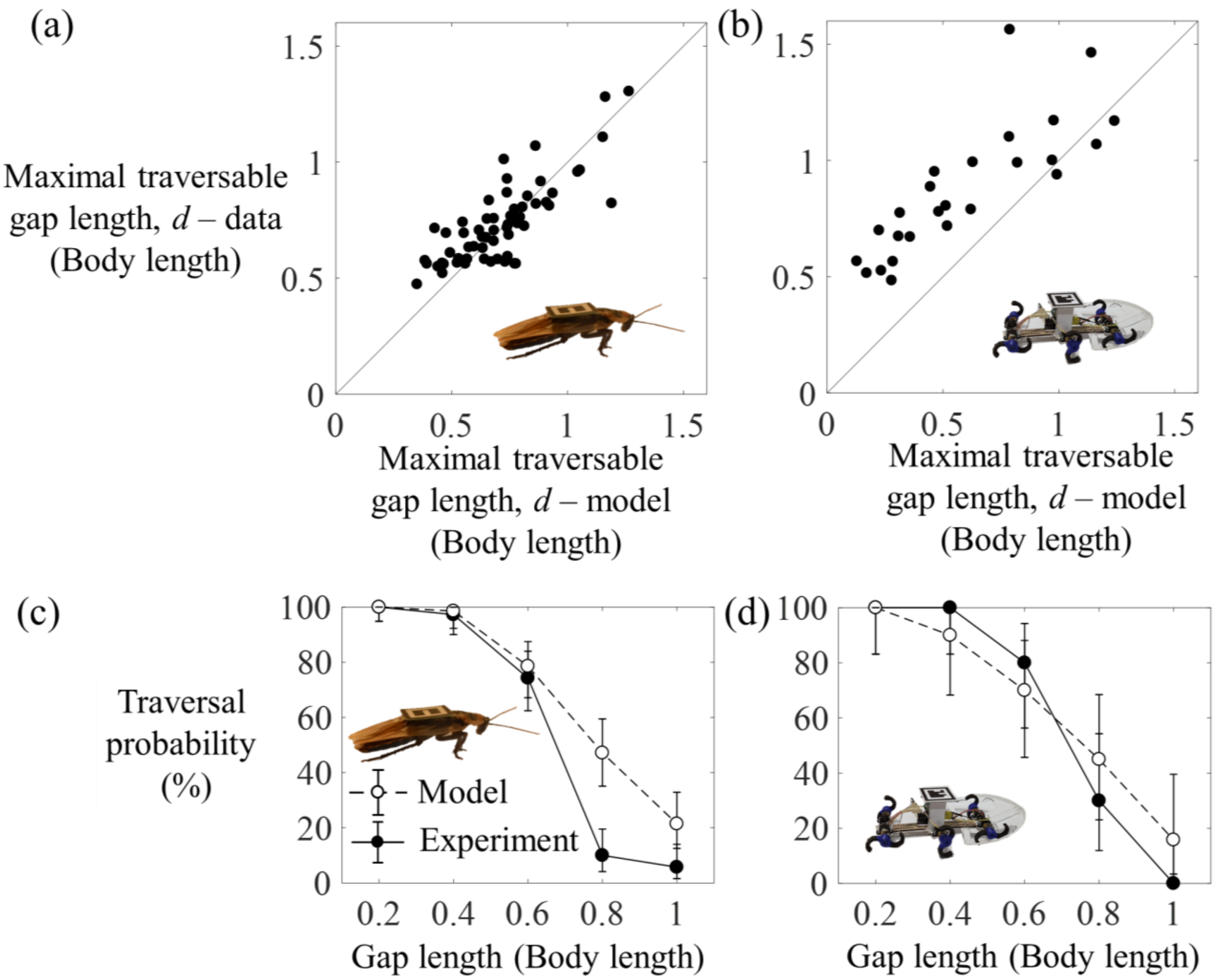

**Figure 5.** The template well predicted dynamic gap traversal. (a, b) Comparison of maximal traversable gap length measured from the infinite length gap experiment with that predicted from the model. The black diagonal line indicates perfect match between data and model. (c, d), Comparison of experimental (filled circles, solid curve) and model-predicted (open circles, dashed curve) traversal probability. Error bars represent 95% confidence interval.

### 3.4. Principles of dynamic gap traversal from template

Our experimentally validated template allowed us to gain insights into general principles of dynamic gap traversal. Using the template, we numerically calculated how maximal traversable gap length depended on approach speed $v_0$, initial body pitch $\beta_0$, and initial body pitch angular velocity $\omega_0$ over a broad range of parameter space. We discovered that maximal traversable gap length increased with all these locomotor parameters (figure 6). This is intuitive because faster running allows an animal or robot to travel forward by a larger distance before its head falls below the surface level, and higher initial body pitch and higher initial pitch angular velocity gave the animal or robot longer time to travel forward before the head fell below surface level.

These general principles from the template gave us two predictions: First, for a given gap length, the animal or robot running at higher approach speed, higher initial body pitch, and/or higher initial body pitch angular velocity should be more likely to successfully traverse. Second, as gap length increased, the approach speed, initial body pitch, and initial body pitch angular velocity required to traverse should also increase (supplementary video 4). Indeed, in our finite length gap experiments, we observed that for all but the smallest gap lengths tested, both the animal and robot ran at higher average approach speeds (measured in Froude number, $Fr$) and had a higher initial body pitch angular velocity when they traversed the gap compared to when they failed (figure 6, circles; figure A2; animal: $P < 0.05$, multiple logistic regression; robot: $P < 0.05$, multiple logistic regression). However, for both the animal and the robot, initial body pitch did not differ between successful and failed trials for all the gap lengths tested ($P > 0.05$, multiple logistic regression). This was likely a result of the small range and large variation of initial body pitch attainable by the animal ($12° \pm 4°$) and the robot ($1° \pm 3°$) when they ran at the speeds in our experiments.

In the infinite length gap experiments, both the animal and robot reached a maximal traversable gap length longer than its body length (figure 5(a, b); animal: 1.25 body length; robot: 1.6 body length). However, traversal of these gap lengths are unlikely. The animal would collide with the far side of the gap at high speeds, making gripping difficult due to the slow reaction time. Additionally, the animal and the robot often could not grip before falling into the gap. Therefore, the model likely predicted a larger maximal traversable gap length than is physically possible due to the grip failure (see section 3.6).

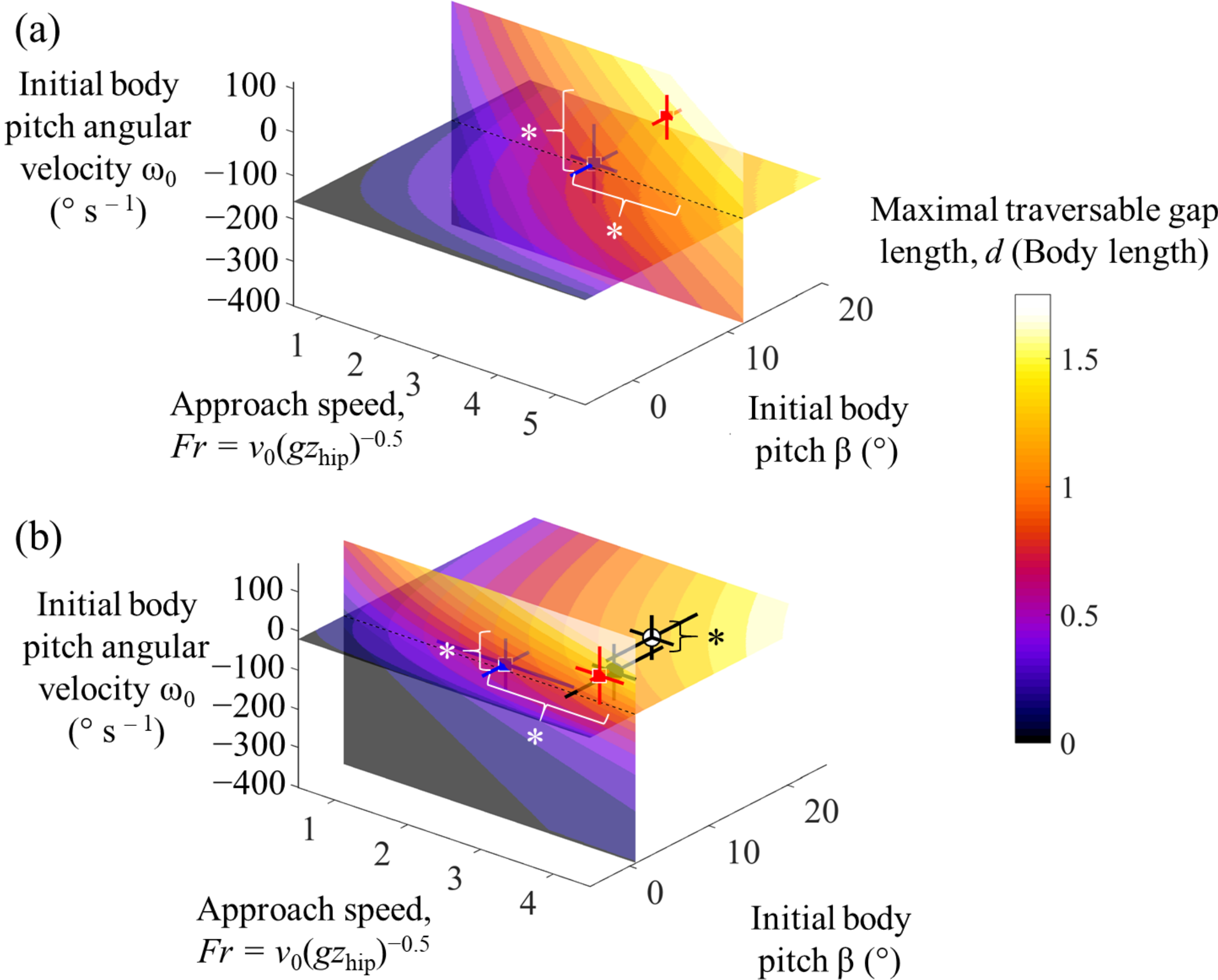

**Figure 6.** The template quantitatively predicted maximal traversable gap length for both the animal and the robot. Maximal traversable gap length as a function of approach speed (shown in Froude number $Fr = v_0(gz_{hip})^{-1/2}$) and initial body pitch, and initial body pitch angular velocity for the animal (a) and robot (b). The surface colors show the model predictions. To create the contour slices, we used the initial body pitch angular velocity (horizontal plane) and the initial body pitch (vertical plane) averaged for all trials for each individual animal and for all robot trials. Circles show the mean ± s.d. for successful (red) and failed (blue) traversal of the 1 body length gap. For the tailed robotfilled circles show the mean ± s.d. without tail activation and open circles show the mean ± s.d. with tail activation (see section 3.5). Bracket and asterisk indicate a statistically significant difference.

Given its simplicity and predictive power of dynamic gap traversal over a broad range of locomotor and terrain parameter space, our model provided the first template (figure 7) for dynamic locomotion in complex 3-D terrains. In addition, because the model did not include any information of legs, we expect that the template can be applied to other types of locomotors, such as wheeled and tracked vehicles [84, 85] and limbless animals [86] and robots [87–89] during high-speed, dynamic gap traversal.

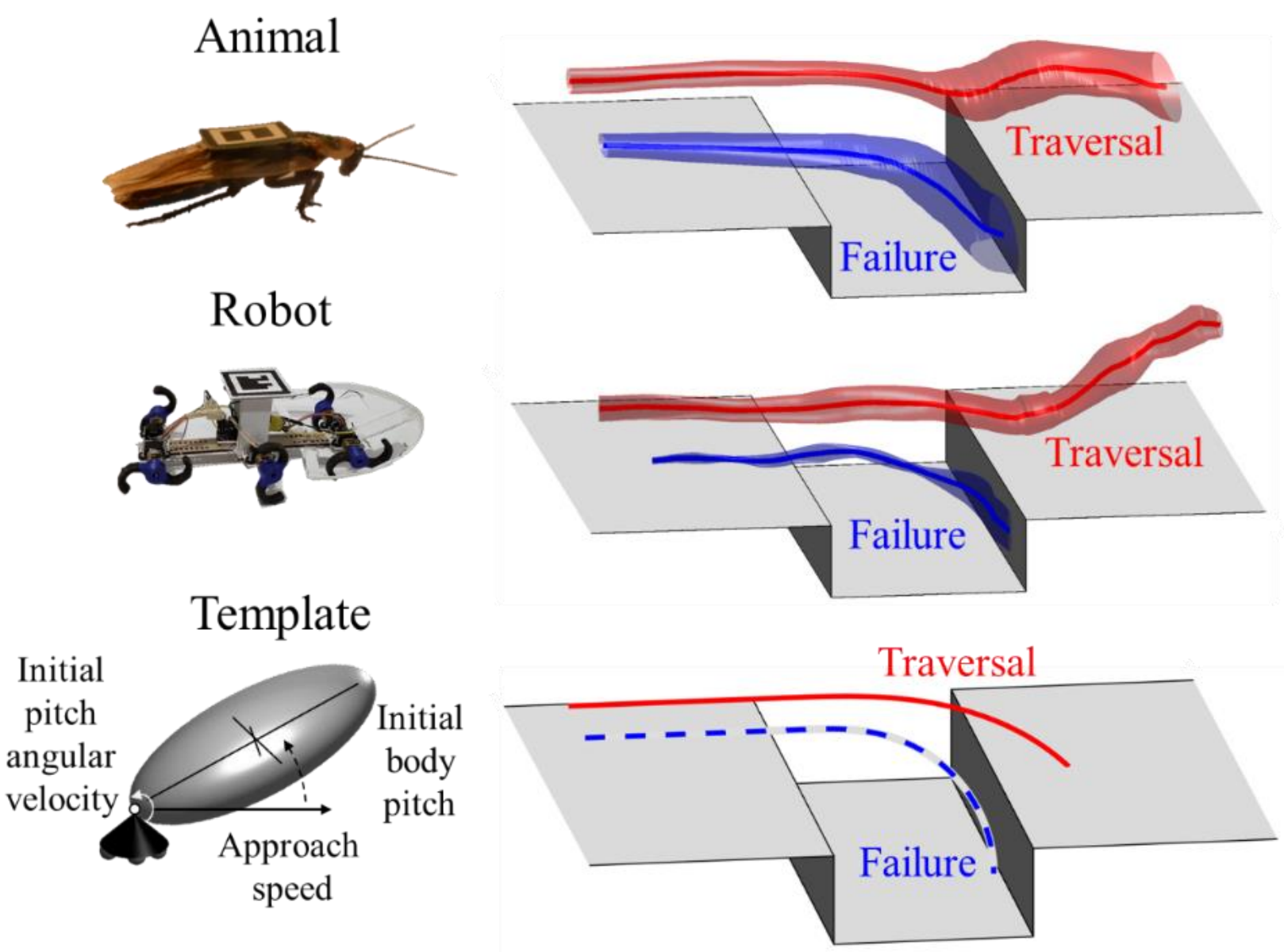

**Figure 7.** Summary of animal, robot, and template body dynamics and body-terrain interaction. Red and blue tubes and curves represent means ± 1 s.d. of vertical and lateral position of the head as the body moved forward towards the gap. The trajectories of successful traversal and failure are slightly offset in the *y*-direction for clarity. For simplicity, only movement perpendicular to the gap (within the *x*-*z* plane) is shown.

### 3.5. Body pitch control to enhance dynamic gap traversal

The principles from our template provided useful control strategies for robots to traverse large gap obstacles. For example, for a given robot already reaching its maximal speed with a head-on approach (zero angle of incidence $\theta_0$), higher initial body pitch and higher initial body pitch angular velocity would allow it to traverse a larger gap. To demonstrate this, we tested the robot with a bio-inspired active tail [21, 33] that enabled adjustment of body pitching (figure 8(a)). As the robot approached the gap, the tail was rotated backwards and suddenly stopped, and its angular momentum was quickly transferred to the body, causing the robot to pitch upward (supplementary video 3).

As predicted by our template, we found that the active tail significantly increased the robot's ability to traverse a large gap by increasing $\omega_0$ and the robot's pitch moment of inertia (figure 9). At an approach speed of 190 cm $s^{-1}$ ± 20 cm $s^{-1}$, the robot's initial body pitch angular velocity increased from –11 ± 74 ° $s^{-1}$ without tail actuation to 57 ± 57 ° $s^{-1}$ with tail actuation (figure 6(d), circles; $P = 0.011$, ANOVA). Although tail actuation did not significantly increase initial body pitch ($P = 0.8$, Student's *t*-test), it did increase the maximal body pitch as the robot ran over the near edge of the gap, from 5° ± 6° without tail actuation to 8° ± 6° with tail actuation ($P = 0.045$, ANOVA). Together, these changes in body pitching not only increased the robot's traversal probability for all but the smallest gap tested, but also allowed it to traverse a gap as large as 1.2 body length, a 50% increase (figure 9(b), open circles).

Although we only demonstrated using an active tail to increase initial body pitch and initial body pitch angular velocity to aid large gap obstacle traversal, body pitch can continue to be controlled during the entire falling phase [21, 74]. Further, other body pitch control methods may also be employed, such as pushing more forcefully with fore and mid legs [34, 36] and hyperextending a flexible body [90]. Future studies should test the feasibility and performance of these control strategies and add sensory feedback [2] to further improve robot performance traversing large gap obstacles.

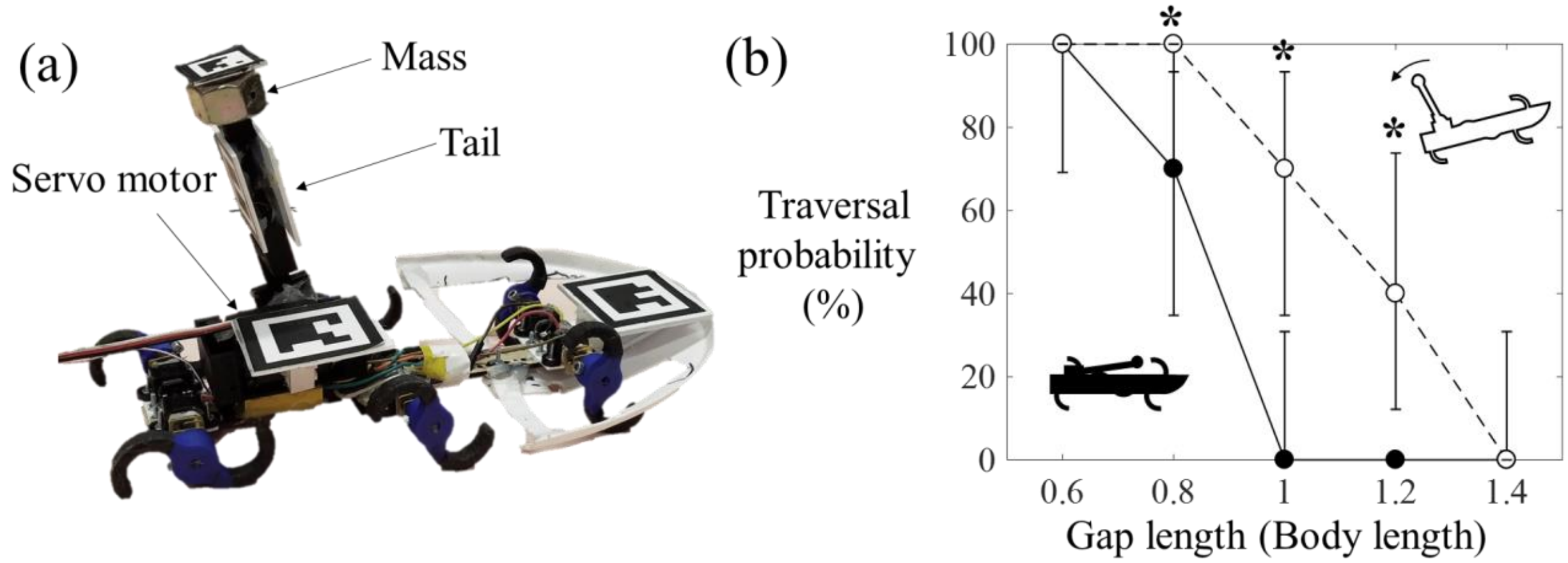

**Figure 8.** Tailed robot experiments to demonstrate the novel control strategy for dynamic gap traversal. (a) The robot with an active tail. (b) Gap traversal probability for the tailed robot with (open circles) and without (filled circles) tail actuation while running at the same speed of 190 ± 20 cm $s^{-1}$.

### 3.6. Body flexibility and leg gripping ability enhance large gap traversal

Our template of dynamic large gap traversal well captured body dynamics until the head bridged across the gap using kinetic energy. However, for traversal, the animal or robot must also be able to continue to move forward until its entire body made it across the gap. This is especially challenging over a large gap, because when the head bridged, the majority of the body or even the entire body is still above the gap without ground reaction forces (figure 3(a, b, c, d), frame 2). Therefore, generating sufficient upward and forward forces against the far edge of the gap is essential for traversal.

Careful examination revealed that successful traversal of a large gap obstacle after gap bridging by the head often required additional leg gripping and pulling (66% and 100% of the time for 0.8 and 1 body length gaps, respectively), presumably using sensory feedback (figure 9(a)). The animal's ability to flex its body and use the various structures on its articulated legs and feet [91–93] to grip when they touched the far side surface played an important role in this process. We observed that, after the animal's head bridged across a large gap, the body often flexed substantially, while its middle and fore legs pulled on the far edge and its hind legs pushed against the vertical surface (figure 9(a), frames 3, 4, 5). Body flexion not only allowed fore legs to better reach forward and downward to pull but also allowed hind legs to reach downward and forward to push [90]. When the animal was unable to do these sufficiently in time, its body pitched backwards, resulting in falling back into the gap even when the head bridged across (figure 9(b), frames 3, 4, and 5).

By contrast, the robot the robot's body and legs are relatively rigid and lacked gripping mechanisms to pull itself onto the far side of gap (figure 9(c)). Even when the robot succeeded in traversing a large gap, it did not grip, but simply continued to use the same open-loop gait. As a result, even when the head bridged across the gap, the robot was more likely (19% probability) to fall backwards and fall back into the gap than the animal (3% probability) ($P < 0.0001$, repeated-measures ANOVA). In addition, the probability of grip failure for both the animal and robot increased for the largest gap lengths (animal: $P < 0.0001$, repeated-measures ANOVA; robot: $P < 0.0001$, ANOVA).

Based on these observations, we posit that a two-link body modeling the flexible body with three spring legs modeling fore, mid, and hind feet is a likely candidate for an anchor-level model [37] to better capture the dynamics of the final phase of dynamic traversal of large gaps and to further explore the role of active body pitch control [21] (figure 9(d)). Future experiments should better understand the biological principles of such active body and leg use and validate the anchor model, and use them to improve the ability of robots to dynamically traverse large gaps.

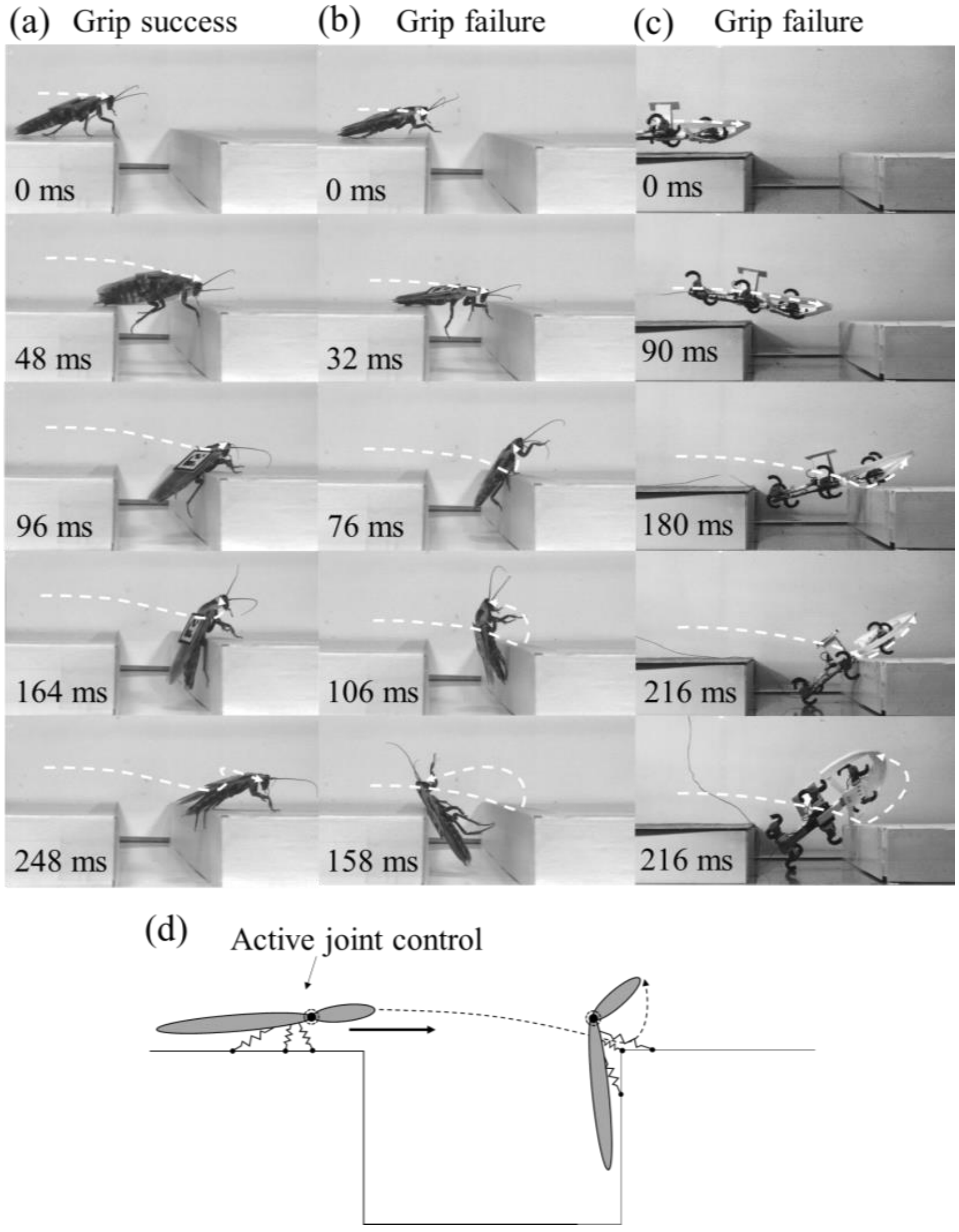

**Figure 9.** Traversal often required gripping and body flexion. (a) Successful leg gripping aided by body flexion by the animal when dynamically traversing a large gap. (b) Grip failure of the animal. (c) Grip failure of the robot. (d) A proposed two-link anchor model for dynamic gap traversal with active body flexion and leg gripping.

# 4. Conclusions

We comparatively studied rapid-running discoid cockroaches and a cockroach-inspired robot as a physical model to discover the performance limits and general principles of dynamic traversal of a large gap obstacle. We discovered that similar to bridging small gaps between footholds on low contact area surfaces [28] and uneven terrain [6], both the animal and the robot can use translational and rotational kinetic energy to dynamically traverse a gap obstacle as large as its body length. This is rarely possible during quasi-static gap traversal [4, 38, 94] and more similar to snakes using kinetic energy to lunge across large gaps [19]. Traversal was less likely as the gap became larger and was well predicted by whether the head bridged across the gap. Inspired by the similarity in animal and robot observations, we created a template that well described body dynamics during passive falling over the gap and quantitatively predicted traversal performance. Our template revealed that high approach speed, high initial high body pitch, and high initial body pitch angular velocity all facilitated dynamic traversal, by allowing the head to travel further to bridge a gap larger than would be possible during quasi-static gap traversal [4, 38, 94]. Despite their similarities, the animal is still far better than the robot at dynamically traversing large gap obstacles, thanks to its stronger ability to grip, push, and pull itself onto the far side of the gap.

Our study is a major step in expanding the emerging field of terradynamics of biological and robotic locomotion [8]. Our template is the first to quantitatively predict dynamic locomotion beyond planar surfaces [10, 37, 43] and expanded the usefulness of templates [37] into complex 3-D terrains. Future studies to systematically vary locomotor and terrain parameters [8, 72, 95, 96] and create new templates for other types of terrains will advance our understanding of how animals move in nature and improve robotic locomotion in complex natural and artificial environments. An immediate next step is to discover general principles for dynamic traversal of another simple yet general obstacle, a large bump, which we explore in our companion study [67].

## Acknowledgements

We thank Yuanfeng Han, Tom Libby, Simon Sponberg, Bob Full, and two anonymous reviewers for helpful suggestions; Nastasia Winney and Rafael de la Tijera Obert for help with preliminary experiments; Yuanfeng Han for help with experimental setup; and Nicole Mazouch and Nate Hunt for providing animals during preliminary experiments. This work is funded by a Burroughs Welcome Fund Career Award at the Scientific Interface, an Army Research Office Young Investigator Award, and The Johns Hopkins University Whiting School of Engineering start-up funds to C L. Author contributions: S W G designed study, performed experiments, analyzed data, developed template, and wrote the paper; C Y developed robot tail mechanism and assisted with robot experiments; R O and Z R developed robot; C L designed and supervised study, assisted with template development, and wrote the paper.

## References

[1] Kaspari, M , Weiser M D 2016 The Size-Grain Hypothesis and Interspecific Scaling in Ants **13** 530–8

[2] Dickinson M H, Farley C T, Full R J, Koehl M A R, Kram R and Lehman S 2000 How Animals Move : An Integrative View **288** 100–7

[3] Watson J T, Ritzmann R E, Zill S N and Pollack A J 2002 Control of obstacle climbing in the cockroach, Blaberus discoidalis. I. Kinematics *J. Comp. Physiol. A* **188** 39–53

[4] Blaesing B and Cruse H 2004 Stick insect locomotion in a complex environment: climbing over large gaps *J. Exp. Biol.* **207** 1273–86

[5] Kohlsdorf T and Biewener A A 2006 Negotiating obstacles: Running kinematics of the lizard Sceloporus malachiticus *J. Zool.* **270** 359–71

[6] Sponberg S and Full R J 2008 Neuromechanical response of musculo-skeletal structures in cockroaches during rapid running on rough terrain. *J. Exp. Biol.* **211** 433–46

[7] Jayne B C and Riley M a 2007 Scaling of the axial morphology and gap-bridging ability of the brown tree snake, Boiga irregularis. *J. Exp. Biol.* **210** 1148–60

[8] Li C, Pullin A O, Haldane D W, Lam H K, Fearing R S and Full R J 2015 Terradynamically streamlined shapes in animals and robots enhance traversability through densely cluttered terrain

*Bioinspir. Biomim* **10**

[9] Alexander R M 2003 *Principles of animal locomotion* (Princeton University Press)

[10] Goldman D I, Chen T S, Dudek D M and Full R J 2006 Dynamics of rapid vertical climbing in cockroaches reveals a template. *J. Exp. Biol.* **209** 2990–3000

[11] Theunissen L M and Durr V 2013 Insects use two distinct classes of steps during unrestrained locomotion *PLoS One* **8** 1–18

[12] Higham T E, Korchari P and Mcbrayer L D 2011 How to climb a tree: Lizards accelerate faster, but pause more, when escaping on vertical surfaces *Biol. J. Linn. Soc.* **102** 83–90

[13] Byrnes G and Jayne B C 2014 Gripping during climbing of arboreal snakes may be safe but not economical. *Biol. Lett.* **10**20140434-

[14] Byrnes G and Jayne B C 2010 Substrate diameter and compliance affect the gripping strategies and locomotor mode of climbing boa constrictors. *J. Exp. Biol.* **213** 4249–56

[15] Autumn K, Hsieh S T, Dudek D M, Chen J, Chitaphan C and Full R J 2006 Dynamics of geckos running vertically. *J. Exp. Biol.* **209** 260–72

[16] Ritzmann R E, Pollack A J, Archinal J, Ridgel A L and Quinn R D 2005 Descending control of body attitude in the cockroach Blaberus discoidalis and its role in incline climbing *J. Comp. Physiol. A* **191** 253–64

[17] Mongeau J, Mcrae B, Jusufi A, Birkmeyer P, Hoover A M, Fearing R and Full R J 2012 Rapid Inversion : Running Animals and Robots Swing like a Pendulum under Ledges *PLoS One* **7**

[18] Cruse H 1976 The Control of Body Position in the Stick Insect (Carausius morosus), when Walking over Uneven Surfaces *Biol. Cybern.* **33** 25–33

[19] Byrnes G and Jayne B C 2012 The effects of three-dimensional gap orientation on bridging performance and behavior of brown tree snakes (Boiga irregularis) *J. Exp. Biol.* **215** 2611–20

[20] Hoefer K M and Jayne B C 2013 Three-Dimensional Locations of Destinations Have Species-Dependent Effects on the Choice of Paths and the Gap-Bridging Performance of Arboreal Snakes *J. Exp. Zool. Part A* **319** 124–37

[21] Libby T, Moore T Y, Chang-Siu E, Li D, Cohen D J, Jusufi A and Full R J 2012 Tail-assisted pitch control in lizards, robots and dinosaurs *Nature* **481** 181–4

[22] Watson J T, Ritzmann R E and Pollack A J 2002 Control of climbing behavior in the cockroach, Blaberus discoidalis. II. Motor activities associated with joint movement *J. Comp. Physiol. A* **188** 55–69

[23] Harley C M, English B a and Ritzmann R E 2009 Characterization of obstacle negotiation behaviors in the cockroach, Blaberus discoidalis. *J. Exp. Biol.* **212** 1463–76

[24] Olberding J P, McBrayer L D and Higham T E 2012 Performance and three-dimensional kinematics of bipedal lizards during obstacle negotiation *J. Exp. Biol.* **215** 247–55

[25] Parker S E and McBrayer L D 2016 The effects of multiple obstacles on the locomotor behavior and performance of a terrestrial lizard *J. Exp. Biol.* **219** 1004–13

[26] Zurek D B and Gilbert C 2014 Static antennae act as locomotory guides that compensate for visual motion blur in a diurnal , keen-eyed predator *Proc. R. Soc. B* **281** 20133072

[27] Birn-Jeffery a. V. and Daley M 2012 Birds achieve high robustness in uneven terrain through active control of landing conditions *J. Exp. Biol.* **215** 2117–27

[28] Spagna J C, Goldman D I, Lin P-C, Koditschek D E and Full R J 2007 Distributed mechanical feedback in arthropods and robots simplifies control of rapid running on challenging terrain. *Bioinspir. Biomim.* **2** 9–18

[29] Daley M and Biewener A A 2006 Running over rough terrain reveals limb control for intrinsic stability. *Proc. Natl. Acad. Sci. U. S. A.* **103** 15681–6

[30] Collins C E, Self J D, Anderson R A and McBrayer L D 2013 Rock-dwelling lizards exhibit less sensitivity of sprint speed to increases in substrate rugosity *Zoology* **116** 151–8

[31] Jayaram K and Full R J 2016 Cockroaches traverse crevices, crawl rapidly in confined spaces, and inspire a soft, legged robot *Proc. Natl. Acad. Sci.* **113** 950–7

[32] Altendorfer R, Moore N, Komsuoglu H, Buehler M, Brown H B, Mcmordie D, Saranli U, Full R and Koditschek D E 2001 RHex: A biologically inspired hexapod runner *Auton. Robots* **11** 207–13

[33] Brill A L, De A, Johnson A M and Koditschek D E 2015 Tail-assisted rigid and compliant legged leaping *IEEE Int. Conf. Intell. Robot. Syst.* 6304–11

[34] Johnson A M and Koditschek D E 2013 Toward a vocabulary of legged leaping *Proc. - IEEE Int. Conf. Robot. Autom.* 2568–75

[35] Haynes G C, Pusey J, Knopf R, Johnson A and Koditschek D 2012 Laboratory on legs: an architecture for adjustable morphology with legged robots *SPIE Defense, Secur. Sens.* **8387**83870–83870W

[36] Chou Y C, Huang K J, Yu W S and Lin P C 2015 Model-based development of leaping in a hexapod robot *IEEE Trans. Robot.* **31** 40–54

[37] Full R J and Koditschek D E 1999 Templates and anchors: neuromechanical hypotheses of legged locomotion on land *J. Exp. Biol.* **2** 3–125

[38] Dürr V 2001 Stereotypic leg searching movements in the stick insect: kinematic analysis, behavioural context and simulation. *J. Exp. Biol.* **204** 1589–604

[39] Cowan N J and Full R J 2006 Task-level control of rapid wall following in the American cockroach *J. Exp. Biol.* **209** 3043–3043

[40] Okada J and Toh Y 2004 Spatio-temporal patterns of antennal movements in the searching cockroach. *J. Exp. Biol.* **207** 3693–706

[41] Dean J and Wendler G 1983 Stick Insect Locomotion on a Walking Wheel: Interleg Coordination of Leg Position *J. Exp. Biol.* **103** 75–94

[42] Theunissen L M, Vikram S and Dürr V 2014 Spatial coordination of foot contacts in unrestrained climbing insects. *J. Exp. Biol.* **217** 3242–53

[43] Lee J, Sponberg S, Loh O, Lamperski A, Full R and Cowan N 2007 Templates and Anchors for Antenna-Based Wall Following in Cockroaches and Robots *IEEE Trans Robot* **24** 1–14

[44] Okada J and Toh Y 2006 Active tactile sensing for localization of objects by the cockroach antenna *J. Comp. Physiol. A* **192** 715–26

[45] Ritzmann R E, Harley C M, Daltorio K A, Tietz B R, Pollack A J, Bender J A, Guo P,

Horomanski A L, Kathman N D, Nieuwoudt C, Brown A E and Quinn R D 2012 Deciding which way to go: How do insects alter movements to negotiate barriers *Front. Neurosci.* **6** 1–10

[46] Okada J and Toh Y 2000 The role of antennal hair plates in object-guided tactile orientation of the cockroach (Periplaneta americana) *J. Comp. Physiol. A* **186** 849–57

[47] Daltorio K A, Mirletz B T, Sterenstein A, Cheng J C, Watson A, Kesavan M, Bender J A, Ritzmann R E and Quinn R D 2014 How Cockroaches Employ Wall-Following for Exploration 72–83

[48] Tucker D B and Mcbrayer L D 2012 Overcoming obstacles: The effect of obstacles on locomotor performance and behaviour *Biol. J. Linn. Soc.* **107** 813–23

[49] More H L, Hutchinson J R, Collins D F, Weber D J, Aung S K H and Donelan J M 2010 Scaling of sensorimotor control in terrestrial mammals. *Proc. Biol. Sci.* **277** 3563–8

[50] Jayaram K, Mongeau J, Mcrae B and Full R J 2010 High-speed horizontal to vertical transistions in running cockroaches reveals a priciple of robustness *Integr. Comp. Biol.* E83

[51] Kubow T M 1999 The role of the mechanical system in control: a hypothesis of self-stabilization in hexapedal runners *Philos. Trans. R. Soc. B Biol. Sci.* **354** 849–61

[52] Jindrich D L and Full R J 2002 Dynamic stabilization of rapid hexapedal locomotion. *J. Exp. Biol.* **205** 2803–23

[53] Dudek D M and Full R J 2006 Passive mechanical properties of legs from running insects. *J. Exp. Biol.* **209** 1502–15

[54] Khatib O 1986 Real-Time Obstacle Avoidance for Manipulators and Mobile Robots *Int. J. Rob. Res.* **5** 90–8

[55] Rimon E and Koditschek D E 1992 Exact Robot Navigation using artificial Potential Functions *IEEE Trans. Robot. Autom.* **8** 501–18

[56] Leonard J J and Durrant-Whyte H F 1991 Simultaneous map building and localization for an autonomous mobile robot *Intell. Robot. Syst.* **3** 1442–7

[57] Koditschek D E 1987 Exact robot navigation by means of potential functions: Some topological considerations *IEEE Int. Conf. Robot. Autom.* **4** 1–6

[58] Thrun S, Burgard W and Fox D 2000 A real-time algorithm for mobile robot mapping with applications to\nmulti-robot and 3D mapping *IEEE Int. Conf. Robot. Autom.* **1** 321–8

[59] Birkmeyer P, Peterson K and Fearing R S 2009 DASH: A dynamic 16g hexapedal robot *IEEE/RSJ Int. Conf. Intell. Robot. Syst.* 2683–9

[60] Hoover A M, Burden S, Fu X Y, Sastry S S and Fearing R S 2010 Bio-inspired design and dynamic maneuverability of a minimally actuated six-legged robot *IEEE/RSJ Int. Conf. Intell. Robot. Syst.* 869–76

[61] Lewinger W a., Harley C M, Watson M S, Branicky M S, Ritzmann R E and Quinn R D 2009 Animal-inspired sensing for autonomously climbing or avoiding obstacles *Appl. Bionics Biomech.* **6** 43–61

[62] Schmitt J and Holmes P 2000 Mechanical models for insect locomotion: Dynamics and stability in the horizontal plane I: Theory *Biol. Cybern.* **83** 501–27

[63] Schmitt J and Holmes P 2000 Mechanical models for insect locomotion: dynamics and stability in the horizontal plane II. Application. *Biol. Cybern.* **83** 517–27

[64] Boggess M J, Schroer R T, Quinn R D and Ritzmann R E 2004 Mechanized cockroach footpaths enable cockroach-like mobility *IEEE Int. Conf. Robot. Autom.* **3**

[65] Schmitt J and Holmes P 2001 Mechanical models for insect locomotion: Stability and parameter studies *Phys. D Nonlinear Phenom.* **156** 139–68

[66] Haldane J B S 1973 On Being the Right Size *Annu. Rev. Microbiol.* **27** 119–32

[67] Gart S and Li C 2017 Dynamic locomotion of insects and legged robots over large 3-D obstacles II. Body-terrain interaction affects high bump traversal *Bioinspir. Biomim*

[68] Marvi H and Hu D L 2012 Friction enhancement in concertina locomotion of snakes *J. R. Soc. Interface* **9** 3067–80

[69] Bell W J, Roth L M and Nalepa C A 2007 *Cockroaches. Ecology, Behaviour and Natural History* (Baltimore, MD, USA: The Johns Hopkins University Press)

[70] Aguilar J, Zhang T, Qian F, Kingsbury M, McInroe B, Mazouchova N, Li C, Maladen R, Gong C, Travers M, Hatton R L, Choset H, Umbanhowar P B and Goldman D I 2016 A review on locomotion robophysics: the study of movement at the intersection of robotics, soft matter and dynamical systems *Reports Prog. Phys.* 1–61

[71] Full B Y R J and Tu M S 1990 Mechanics of six-legged runners **148** 129–46

[72] Qian F, Daffon K, Zhang T and Goldman D I 2013 an Automated System for Systematic Testing of Locomotion on Heterogeneous Granular Media *Nature-Inspired Mob. Robot.* 547–54

[73] Holmes P, Full R J, Koditschek D and Guckenheimer J 2006 The Dynamics of Legged Locomotion: Models, Analyses, and Challenges *SIAM Rev.* **48** 207–304

[74] Jusufi a, Kawano D T, Libby T and Full R J 2010 Righting and turning in mid-air using appendage inertia: reptile tails, analytical models and bio-inspired robots. *Bioinspir. Biomim.* **5** 45001

[75] Koditschek D E, Full R J and Buehler M 2004 Mechanical aspects of legged locomotion control *Arthropod Struct. Dev.* **33** 251–72

[76] Crall J D, Gravish N, Mountcastle A M and Combes S A 2015 BEEtag : A Low-Cost , Image-Based Tracking System for the Study of Animal Behavior and Locomotion *PLoS One* **10** e0136487

[77] Ting L H, Blickhan R and Full R J 1994 Dynamic and Static Stability in Hexapedal Runners *J. Exp. Biol.* **197** 251–69

[78] Halloy J, Sempo G, Caprari G, Rivault C, Asadpour M, Tâche F, Saïd I, Durier V, Canonge S, Amé J M, Detrain C, Correll N, Martinoli A, Mondada F, Siegwart R and Deneubourg J L 2007 Social integration of robots into groups of cockroaches to control self-organized choices *Science (80-. ).* **318** 1155–8

[79] Hedrick T L 2008 Software techniques for two- and three-dimensional kinematic measurements of biological and biomimetic systems *Bioinspir. Biomim* **3**

[80] Alexander R M N and Jayes A S 1983 A dynamic similarity hypothesis for the gaits of quadrupedal mammals *J. Zool.* **201** 135–52

[81] Seyfarth A, Geyer H, Günther M and Blickhan R 2002 A movement criterion for running *J. Biomech.* **35** 649–55

[82] Seyfarth A 2003 Swing-leg retraction: a simple control model for stable running *J. Exp. Biol.* **206** 2547–55

[83] Blickhan R 1989 The spring-mass model for running and hopping *J. Biomech.* **22** 1217–27

[84] Bekker M G 1960 *Off-the-road locomotion; research and development in terramechanics* (Ann Arbor, MI: University of Michigan Press)

[85] Meirion-Griffith G and Spenko M 2010 An empirical study of the terramechanics of small unmanned ground vehicles *IEEE Aerosp. Conf.* 1–6

[86] Gray J and Lissmann H W 1950 The kinetics of the locomotion of the grass-snake *J. Exp. Biol.* **26** 354–67

[87] Gonzalez-Gomez J, Zhang H, Boemo E and Zhang J 2006 Locomotion of a Modular Robot with Eight Pitch-Yaw-Connecting Modules *9th Int. Conf. Climbing Walk. Robot.*

[88] Ohno H and Hirose S 2001 Design of slim slime robot and its gait of locomotion *IEEE/RSJ Int. Conf. Intell. Robot. Syst.* **2** 707–15

[89] Tesch M, Lipkin K, Brown I, Hatton R, Peck A, Rembisz J and Choset H 2009 Parameterized and Scripted Gaits for Modular Snake Robots *Adv. Robot.* **23** 1131–58

[90] Boxerbaum A S, Oro J, Peterson G and Quinn R D 2008 The latest generation Whegs robot features a passive-compliant body joint *IEEE/RSJ Int. Conf. Intell. Robot. Syst.* 1636–41

[91] Russell P 1975 A contribution to the functional analysis of the foot of the Tokay, Gekko gecko (Reptilia : Gekkonidae) *J. Zool. Lond.* **176** 437–76

[92] Gorb S N 2002 Structural Design and Biomechanics of Friction-Based Releasable Attachment Devices in Insects *Integr. Comp. Biol.* **42** 1127–39

[93] J B W, Roth L M and Nalepa C A 2007 *Cockroaches. Ecology, Behaviour and Natural History* vol 1

[94] Lillywhite H B, LaFrentz J R, Lin Y C and Tu M C 2000 The Cantilever Abilities of Snakes *J. Herpetol.* **34** 523–8

[95] Li C, Umbanhowar P B, Komsuoglu H, Koditschek D E and Goldman D I 2009 Sensitive dependence of the motion of a legged robot on granular media *Proc. Natl. Acad. Sci.* **106** 9932–9932

[96] Li C, Zhang T and Goldman D I 2013 A Terradynamics of Legged Locomotion on Granular Media *Science (80-. ).* **339** 1408–12

## Supplementary Figures

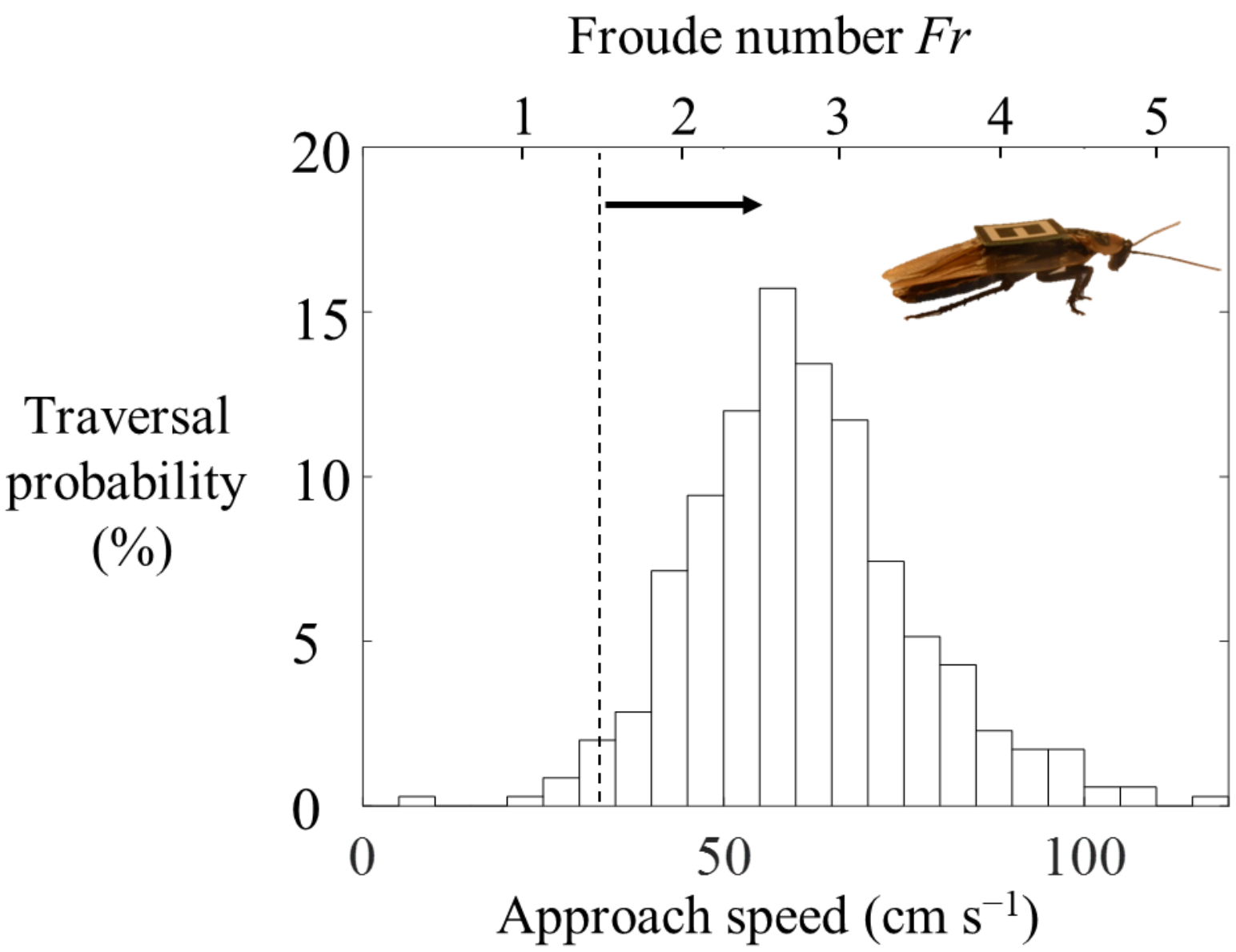

**Figure S1.** Histogram of approach speed for animal finite gap experiments. Vertical dashed line represents an approach speed of 33 cm $s^{-1}$ or $Fr = 1.5$, above which the animal is unlikely able to react to the obstacle in time and likely falls passively (see section 3.1.4).

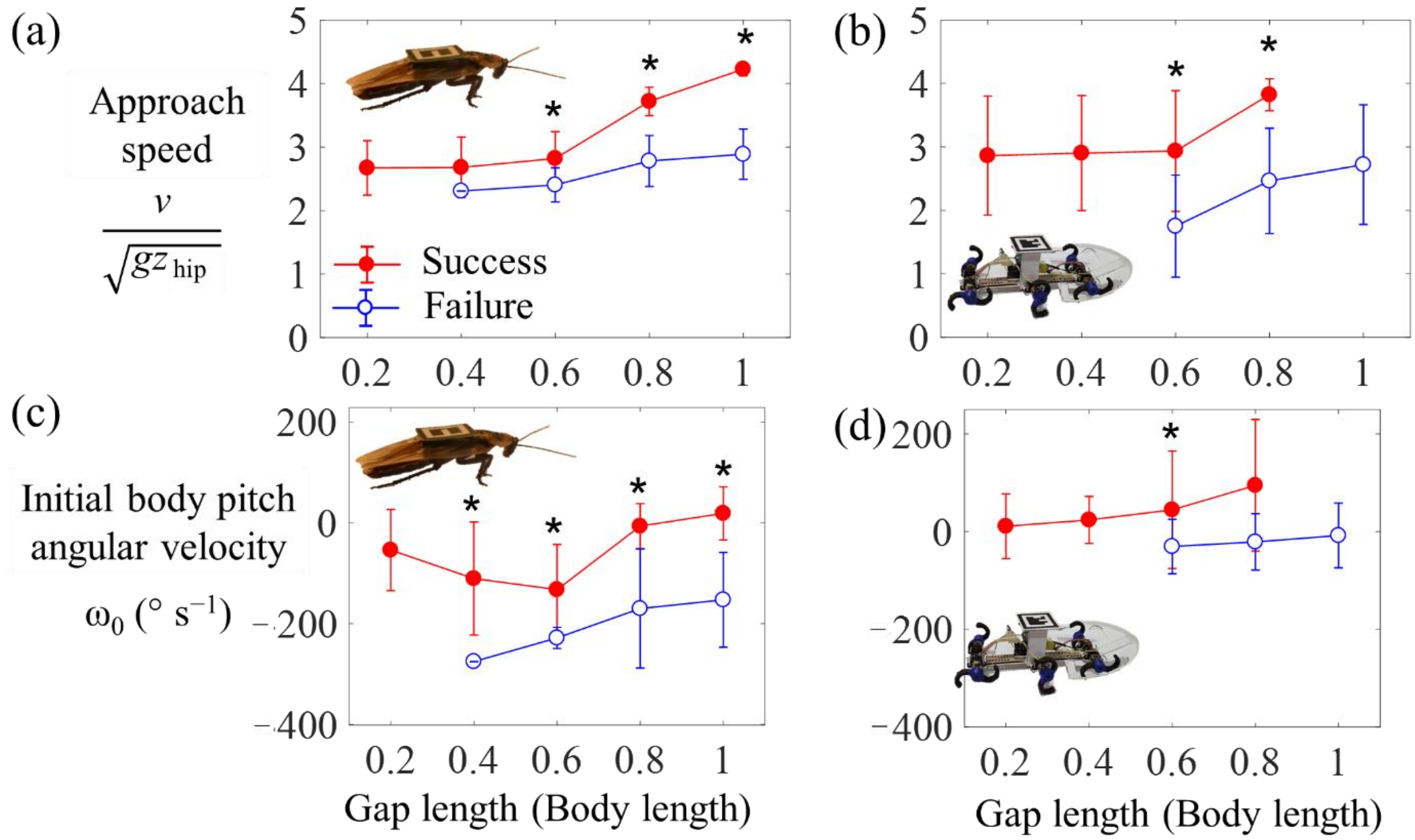

**Figure S2.** Approach speed (a, b) and initial body pitch angular velocity (c, d) is higher during successful traversal than failure for both the animal (a, c) and the robot (b, d). Red filled circles represent successful traversal and blue open circles represent failure to traverse (including grip failure). Error bars represent ± 1 s.d. in (c, d, e, f). Asterisks in indicate a statistically significant difference between the cases of success and failure ($P < 0.05$, multiple logistic regression).

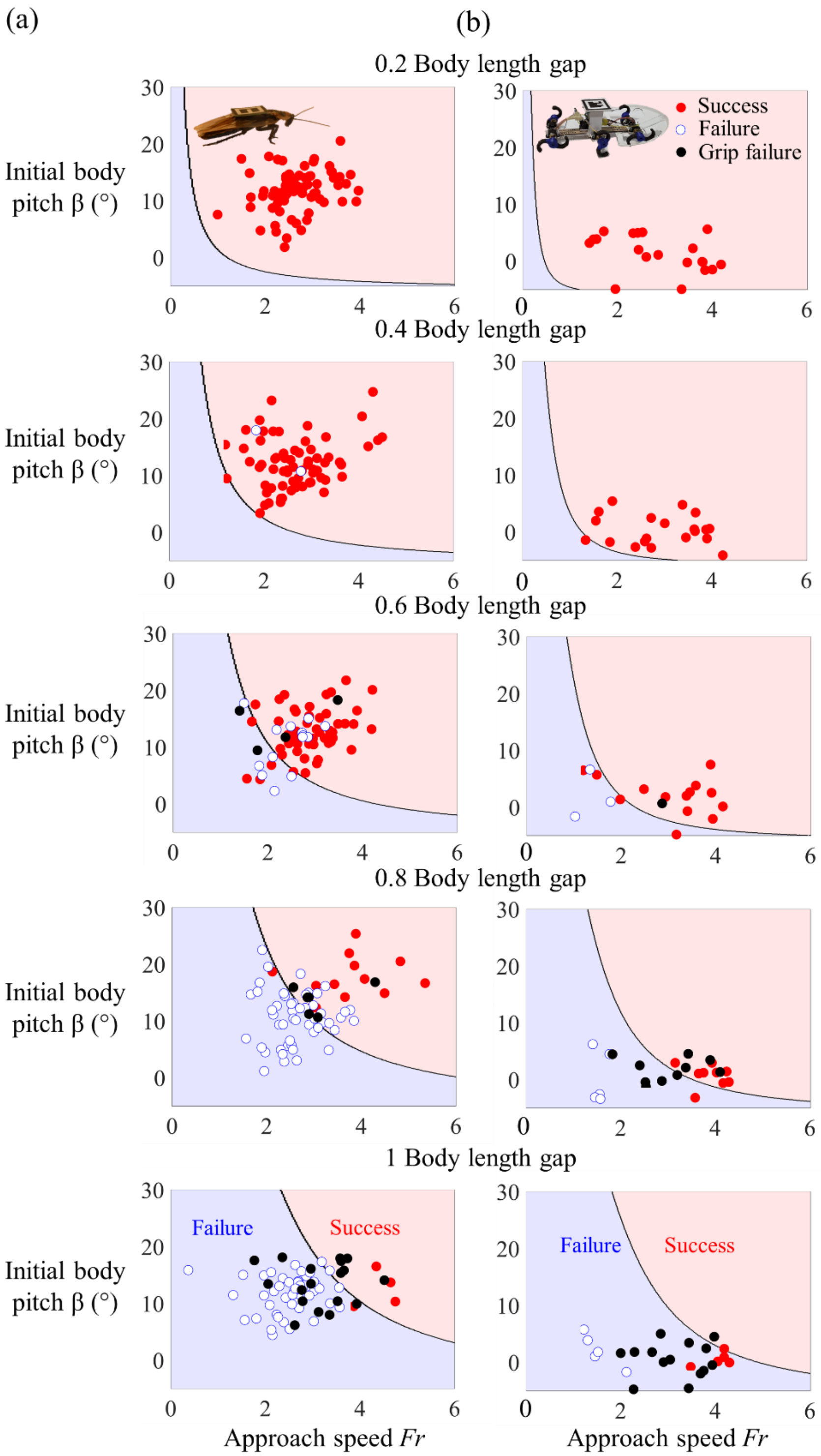

**Figure S3.** Comparison of (a) animal and (b) robot experiments to template predictions of whether each trial results in successful traversal (red) or failure (blue). Circles are experimental data. Red and blue shaded regions are success and failure regimes predicted by the template.

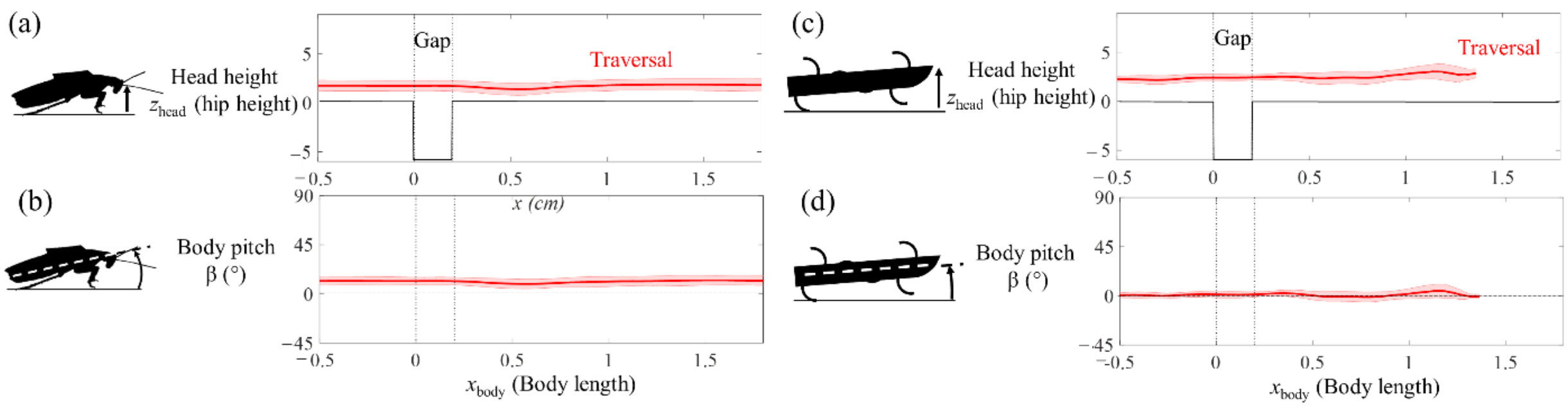

**Figure S4.** Dynamic locomotion of the animal and robot over a 0.2 body length gap obstacle. Representative trials of successful traversal (red solid curves) and failure to bridge the gap (blue dashed curves). (a, c) Head height as a function of forward position of the head. (b, d) Body pitch as a function of forward of the head. Solid red and blue dashed curves and shaded areas represent means ± 1 s.d. for the cases of success and failure. For simplicity, only movement perpendicular to the gap (within the *x*-*z* plane) is shown.

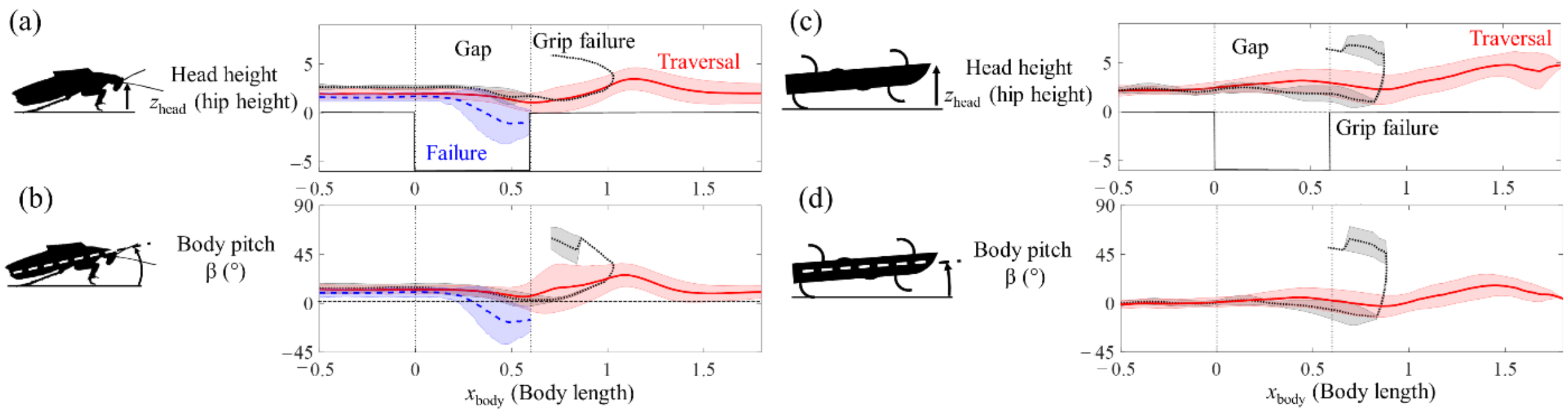

**Figure S5.** Dynamic locomotion of the animal and robot over a 0.4 body length gap obstacle. Representative trials of successful traversal (red solid curves) and failure to bridge the gap (blue dashed curves). (a, c) Head height as a function of forward of the head. (b, d) Body pitch as a function of forward of the head. Solid red and blue dashed curves and shaded areas represent means ± 1 s.d. for the cases of success and failure. For simplicity, only movement perpendicular to the gap (within the *x*-*z* plane) is shown.

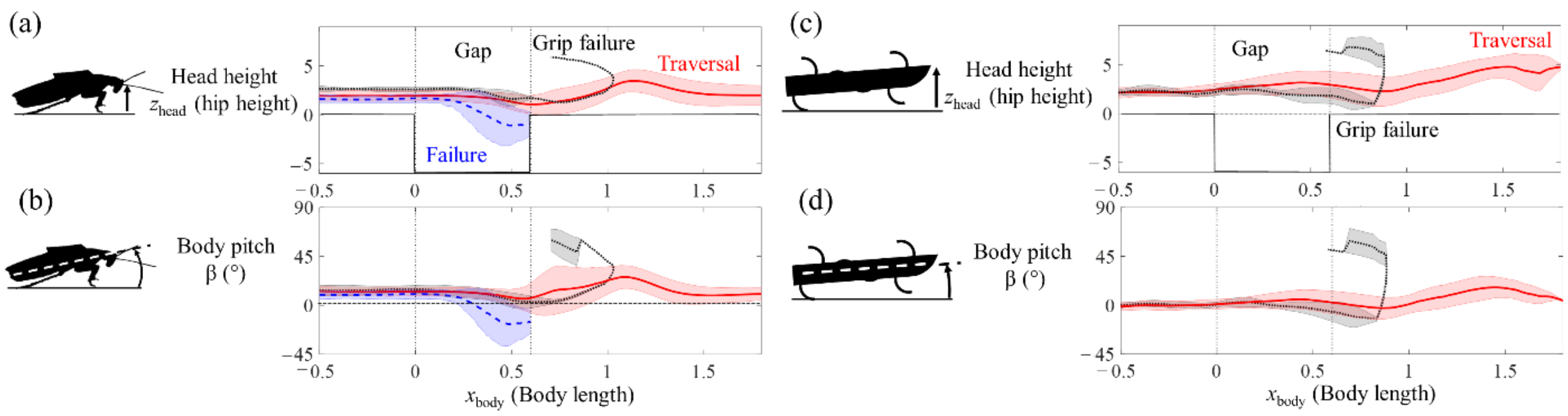

**Figure S6.** Dynamic locomotion of the animal and robot over a 0.6 body length gap obstacle. Representative trials of successful traversal (red solid curves) and failure to bridge the gap (blue dashed curves). (a, c) Head height as a function of forward of the head. (b, d) Body pitch as a function of forward of the head. Solid red and blue dashed curves and shaded areas represent means ± 1 s.d. for the cases of success and failure. For simplicity, only movement perpendicular to the gap (within the *x*-*z* plane) is shown.

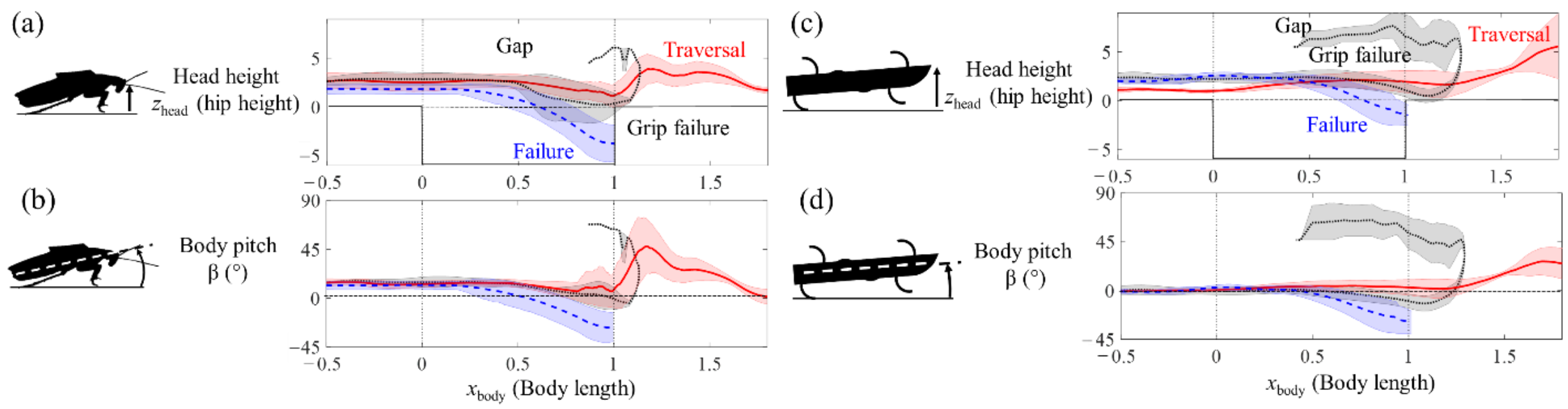

**Figure S7.** Dynamic locomotion of the animal and robot over a 1 body length gap obstacle. Representative trials of successful traversal (red solid curves) and failure to bridge the gap (blue dashed curves). (a, c) Head height as a function of forward of the head. * (b, d) Body pitch as a function of forward of the head. Solid red and blue dashed curves and shaded areas represent means ± 1 s.d. for the cases of success and failure. For simplicity, only movement perpendicular to the gap (within the *x*-*z* plane) is shown.

---

* We noted that when the animal failed to traverse, it often rebounded backwards after impacting the far side of the gap, but we could not track this motion due to the body and gap ledge obscuring the tag. Therefore, we truncated the data at the far edge of the gap.

Supplementary Video 1

https://www.youtube.com/watch?v=woS_BzPiqjc

Supplementary Video 2

https://www.youtube.com/watch?v=UfM-HNBZizM

Supplementary Video 3

https://www.youtube.com/watch?v=yXKVaG9ssiU

Supplementary Video 4

https://www.youtube.com/watch?v=wldsMyd_NMA